 \documentclass[final,5p,times,twocolumn]{elsarticle}

\makeatletter

\@addtoreset{equation}{section}
\makeatother

\usepackage{amsmath,amsfonts,amssymb,amsthm}
\usepackage{mathrsfs}
\usepackage{graphicx}
\usepackage{braket}

\journal{Physica D}

\begin{document}

\begin{frontmatter}



\title{Soliton-phonon scattering problem in 1D nonlinear Schr\"odinger systems with general nonlinearity}


\author{Daisuke A. Takahashi}
\address{Department of Basic Science, The University of Tokyo, Tokyo 153-8902, Japan}
\ead{takahashi@vortex.c.u-tokyo.ac.jp}
\begin{abstract}
	\indent A scattering problem (or more precisely, a transmission-reflection problem) of linearized excitations in the presence of a dark soliton is considered in a one-dimensional nonlinear Schr\"odinger system with a general nonlinearity: $ \mathrm{i}\partial_t \phi = -\partial_x^2 \phi + F(|\phi|^2)\phi $. If the system is interpreted as a Bose-Einstein condensate, the linearized excitation is a Bogoliubov phonon, and the linearized equation is the Bogoliubov equation. We exactly prove that the perfect transmission of the zero-energy phonon is suppressed at a critical state determined by Barashenkov's stability criterion [Phys. Rev. Lett. 77, (1996) 1193.], and near the critical state, the energy-dependence of the reflection coefficient shows a saddle-node type scaling law. The analytical results are well supported by numerical calculation for cubic-quintic nonlinearity. Our result gives an exact example of scaling laws of saddle-node bifurcation in time-reversible Hamiltonian systems. As a by-product of the proof, we also give all exact zero-energy solutions of the Bogoliubov equation and their finite energy extension.

\end{abstract}


\end{frontmatter}


\section{Introduction}
\begin{figure}[t]
		\begin{center}
		\includegraphics[scale=0.9]{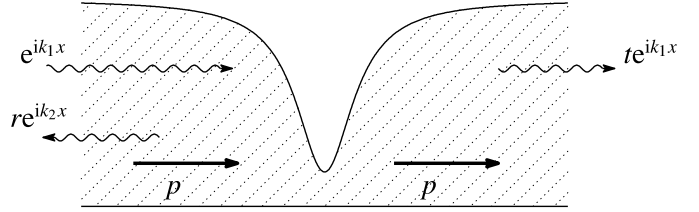}
		\caption{\label{introfigure}A schematic picture of the problem that we consider in this paper.  $ p $ represents a half of the velocity of the dark soliton. The problem is always considered in the comoving frame of the soliton. $ \mathrm{e}^{\mathrm{i}k_1x} $ is an incident wave of a linearized excitation,  $ t\mathrm{e}^{\mathrm{i}k_1x} $ is a transmitted wave, and $ r\mathrm{e}^{\mathrm{i}k_2x} $ is a reflected wave. For more detailed definitions of each quantity, see Sec. \ref{sec:fundamental}.}
		\end{center}
	\end{figure}

	In this paper, we solve a scattering problem of linearized excitations in the presence of a dark soliton in one-dimensional(1D) nonlinear Schr\"odinger (NLS) equation with a general nonlinearity:
	\begin{align}
		\mathrm{i}\partial_t\phi = -\partial_x^2\phi+F(|\phi|^2)\phi, \label{eq:intro001}
	\end{align}
	and discuss the physical and mathematical significance of our results. For a schematic picture, see Fig. \ref{introfigure}. The precise mathematical definition of the problem will be given in Sec. \ref{sec:fundamental}.  If we regard the system as a Bose-Einstein condensate(BEC), the linearized excitation is a Bogoliubov phonon, so the problem can be also called a soliton-phonon scattering problem, as this paper entitled.\\
	\indent  The NLS equation (\ref{eq:intro001}) has a great number of applications in nonlinear optics, superconductors, magnetism, BECs, and so on. Particularly, much attention has been focused on the experimental realizations of BECs in laser-cooled ultracold atoms for more than a decade, because of high-controllability of the system parameters. By using elongated laser beams, low-dimensional systems are realized, and a dark soliton can be created via the phase imprinting method\cite{BurgerPhaseImprinting}. The Bogoliubov theory is also well confirmed\cite{Andrews,Stamper,Steinhauer}. \\
	\indent It is known that 1D NLS with a \textit{cubic} nonlinearity is completely integrable\cite{ZakharovShabat,ZakharovShabat2}. Because of integrability, the linearized equation is also solved exactly\cite{ChenChenHuang,Kovrizhin}, and  the phonon excitations are shown to be completely reflectionless against a soliton for \textit{any} excitation energy. Thus, the problem is trivial in this case. However, when the nonlinear term is generalized, the phonon  has a finite reflection coefficient in general. It is worthy to note that the soliton decay dynamics in the laser-trapped quasi-1D BEC has been well explained by the \textit{quintic} term, which appears as a second-order perturbation of the trapping effect\cite{Muryshev,SinhaChernyKovrizhinBrand,KhaykovichMalomed}, and yields the frictional force between thermal excitation clouds and solitons\cite{Muryshev,SinhaChernyKovrizhinBrand}.  Thus, knowing the scattering properties between solitons and linearized excitations is essential to understand and control the transport of solitons, that is, the transport of stable wave packets. We also mention that the theory of nonpolynomial NLS equation is formulated to describe the confinement effect\cite{SalasnichParolaReatto,MateoDelgado,MateoDelgadoAnnPhys,Salasnich}. The quintic NLS also appears in an effective mean-field description of the Tonks-Girardeau gas\cite{Kolomeisky}. \\
	\indent The NLS equation with an integrability breaking factor is also interesting from the viewpoint of an infinite-dimensional dynamical system and the bifurcation theory. When the potential barrier is added in the cubic NLS equation, there exist stable and unstable stationary supercurrent-flowing solutions\cite{BaratoffBlackburnSchwartz,Hakim}, if the condensate velocity does not exceed a certain critical value. Near the critical point, which separates the stable branch and the unstable branch, it is known that many physical quantities obey saddle-node type scaling, such as an emission period of dark solitons\cite{Hakim,PhamBrachet}, an eigenvalue of a growing mode for the unstable solution\cite{PhamBrachet}, and a transmission coefficient of linearized excitations\cite{Kovrizhin,Kagan,DanshitaYokoshiKurihara}. It is quite nontrivial that the time-reversible Hamiltonian system exhibits the scaling behaviors of saddle-node type, since this bifurcation is normally understood to emerge in time-irreversible phenomena. However, it is not easy to prove these properties analytically or exactly, because of the infinite dimensional character of the system.\\
	\indent On the other hand, as another way to break the integrability, one can consider the generalization of the nonlinearity, that is what we will consider in the present paper.  When the nonlinear term includes a competing interaction, the dark soliton is no longer always stable. One typical example of an unstable dark soliton is a ``bubble'' in a cubic-quintic NLS (CQNLS) system\cite{BarashenkovMakhankov,BarashenkovPanova}. (See also \cite{JYang}.) The most general criterion for the stability of the dark soliton has been shown by Barashenkov\cite{Barashenkov}, and the critical velocity of the soliton is determined by  $ \partial P/\partial v=0 $, where $ v $ is a velocity of the soliton and $ P $ is a renormalized momentum. The existence of the critical velocity lower than the sound velocity and the separation of stable and unstable regions are similar to the phenomena of superflows against a potential barrier. Therefore, we can expect some scaling behavior near the critical state. Furthermore, in the present case, the preserved translational symmetry of the fundamental equation makes it possible to access the problem analytically. \\
	\indent In this paper, we solve the scattering problem of linearized phonon excitations, and exactly show the following: (i) At the critical state determined by Barashenkov's criterion\cite{Barashenkov}, the transparency of the zero-energy phonon is suppressed, and only partial transmission occurs. (ii) Near the critical state, the energy-dependence of the reflection coefficient of low-energy phonons shows saddle-node scaling behavior, regarding the renormalized momentum as a parameter of a normal form of  saddle-node bifurcation.  The obtained analytical results are well confirmed by comparison with the numerical results of CQNLS equation. Our result gives an exact example of scaling laws of saddle-node bifurcation in time-reversible Hamiltonian systems. The proof is based on the exact low-energy expansion of the solution of the linearized equation. Since the exact zero-energy solutions given in this paper are quite general, we believe that our method will also be useful to derive other low-energy physical properties. \\
	\indent The organization of the paper is as follows. In Sec.~\ref{sec:fundamental}, we introduce fundamental equations and see the fundamental properties. The definition of the transmission-reflection problem is also given. In Sec.~\ref{sec:mainresult}, we give a main result and verify it by numerical study of CQNLS equation. Sections~\ref{sec:proof1} and \ref{sec:proof2} are devoted to the proof of main results. Discussions, future perspectives, and conclusions are given in Sec.~\ref{sec:summary}. Some mathematically technical formulae are treated in Appendices.
\section{Fundamental Equations and Definition of the Problem}\label{sec:fundamental}
\subsection{Fundamental equations}
	\indent We begin with the NLS equation with a general nonlinearity
	\begin{align}
		\mathrm{i}\partial_t\phi=-\partial_x^2\phi+F(|\phi|^2)\phi. \label{eq:nls}
	\end{align}
	Here, $ F(\rho) $ is a real-valued function such that $ F(0)=0 $.  The energy functional (Hamiltonian) which yields this equation is
	\begin{align}
		H = \int\!\mathrm{d}x \left(|\partial_x\phi|^2+U(|\phi|^2)\right),
	\end{align}
	where
	\begin{align}
		U(\rho) = \int_0^\rho\mathrm{d}\rho' F(\rho').
	\end{align}
	Letting $ \phi=\phi+\delta\phi $ in Eq. (\ref{eq:nls}), and discarding higher order terms of $ \delta\phi $, one obtains the following linearized equation:
	\begin{align}
		\mathrm{i}\partial_t\delta\phi=\left[-\partial_x^2+ F(|\phi|^2)+|\phi|^2F'(|\phi|^2) \right]\delta\phi+\phi^2F'(|\phi|^2)\delta\phi^*.
	\end{align}
	Writing $ \delta\phi=u $ and $ -\delta\phi^*=v $, one obtains 
	\begin{align}
		\mathrm{i}\partial_t\begin{pmatrix}u\\v\end{pmatrix}=\mathcal{L}\begin{pmatrix}u\\v\end{pmatrix}, \label{eq:tdbogo}
	\end{align}
	where $ \mathcal{L} $ is a  $ 2\times2 $ matrix operator whose components are
	\begin{align}
		\mathcal{L}_{11}^{}&=-\mathcal{L}_{22}^{}=-\partial_x^2+F(|\phi|^2)+|\phi|^2F'(|\phi|^2), \\
		\mathcal{L}_{12}^{}&=-\mathcal{L}_{21}^*=-\phi^2F'(|\phi|^2).
	\end{align}
	We use the notation $ (u,v) $ since it is commonly used by condensed matter physicists. If we interpret the system as BEC, this equation is the Bogoliubov equation which describes the Bogoliubov phonon (or Bogoliubov quasiparticle) \cite{Bogoliubov}. (As a review or a textbook, see, e.g., \cite{FetterWalecka,DalfovoGiorginiPitaevskiiStringari,PethickSmith}.) For this reason, henceforth, we call $ \phi $ the condensate wavefunction or the order parameter, and $ (u,v) $ the Bogoliubov (quasiparticle) wavefunction, though the NLS equation itself has more applications in various fields. \\
	\indent Henceforth we mainly consider the stationary (i.e., time-independent) problem. The stationary NLS equation with chemical potential $ \mu $ is obtained by setting $ \phi(x,t)=\phi(x)\mathrm{e}^{-\mathrm{i}\mu t} $: 
	\begin{align}
		(-\mu-\partial_x^2+F(|\phi|^2))\phi=0. \label{eq:nls2}
	\end{align}
	As will be seen, the value of  $ \mu $ is fixed by the asymptotic form of  $ \phi $ . The stationary Bogoliubov equation with eigenenergy $ \epsilon $ is obtained by setting $ u(x,t)=u(x)\mathrm{e}^{-\mathrm{i}(\epsilon+\mu)t} $ and $  v(x,t)=v(x)\mathrm{e}^{-\mathrm{i}(\epsilon-\mu)t} $:
	\begin{align}
		\epsilon\begin{pmatrix}u\\v\end{pmatrix}=\mathcal{L}_\mu\begin{pmatrix}u\\v\end{pmatrix} \label{eq:bogos}
	\end{align}
	with
	\begin{align}
		\mathcal{L}_\mu:= \mathcal{L}+\begin{pmatrix}-\mu &0 \\ 0& \mu \end{pmatrix}. \label{eq:bogos2}
	\end{align}
\subsection{Bogoliubov phonons in a uniform condensate}\label{subsec:uniform}
	Let us derive the dispersion relation (the energy-momentum relation) of Bogoliubov phonons when the condensate is flowing uniformly: $ \phi(x) = \sqrt{\rho_\infty}\mathrm{e}^{\mathrm{i}(px+\varphi)} $. In order for this $ \phi(x) $ to be the solution of Eq. (\ref{eq:nls2}), the chemical potential must be
	\begin{align}
		\mu=p^2+F(\rho_\infty). \label{eq:cp}
	\end{align}
	The four solutions of Bogoliubov equation (\ref{eq:bogos}) are given by
	\begin{align}
		w_i(x,\varphi):= \begin{pmatrix} \bar{u}_i\, \mathrm{e}^{\mathrm{i}(px+\varphi)} \\ \bar{v}_i\, \mathrm{e}^{-\mathrm{i}(px+\varphi)} \end{pmatrix} \mathrm{e}^{\mathrm{i}k_i x}, \label{eq:bogouniform}
	\end{align}
	where  $ i=1,\,2,\,3, \text{ and }4 $, and the wavenumber $ k_i $s are the roots of the following quartic equation:
	\begin{align}
		(\epsilon-2kp)^2=k^2(k^2+2\rho_\infty F'(\rho_\infty)). \label{eq:uniformdisp}
	\end{align}
	Equation (\ref{eq:uniformdisp}) gives the dispersion relation, and from this dispersion one can see that a half of the Landau's critical velocity (or a half of the sound wave velocity) is given by
	\begin{align}
		p_{\text{L}}=\sqrt{\frac{\rho_\infty F'(\rho_\infty)}{2}}. \label{eq:defofpl}
	\end{align}
	(Note that the sound wave velocity is not $ p_{\text{L}} $ but $ 2p_{\text{L}} $; see the next subsection.) Since $ p_{\text{L}} $ must be real, in order for the uniform condensate to be stable,  $ F'(\rho_\infty)>0 $ must hold. 
	The coefficients $ \bar{u}_i $ and $ \bar{v}_i $ can be, e.g., chosen as follows:
	\begin{align}
		\bar{u}_{i}&=\sqrt{1+\frac{\epsilon-2p k_i^{}+k_i^2}{2p_{\text{L}}^2}}, \label{eq:ucoeff}\\
		\bar{v}_{i}&=\sqrt{1-\frac{\epsilon-2p k_i^{}-k_i^2}{2p_{\text{L}}^2}}. \label{eq:vcoeff}
	\end{align}
	\indent When $ \epsilon>0 $ and $ -p_{\text{L}}<p<p_{\text{L}} $, the quartic equation (\ref{eq:uniformdisp}) has one real positive root, one real negative root, and two complex roots conjugate to each other. We call a real positive (negative) root  $ k_1 \ (k_2) $, and a complex root with positive (negative) imaginary part $ k_3 \ (k_4) $. The low-energy expansions of them are given by
	\begin{align}
		k_1 &= \frac{\epsilon}{2(p+p_{\text{L}})}+O(\epsilon^3), \label{eq:kexpand1} \\
		k_2 &= \frac{\epsilon}{2(p-p_{\text{L}})}+O(\epsilon^3), \label{eq:kexpand2} \\
		k_3 &= 2\mathrm{i}\sqrt{p_{\text{L}}^2-p^2}+\frac{p\epsilon}{2(p_{\text{L}}^2-p^2)}+O(\epsilon^2),  \label{eq:kexpand3} \\
		k_4 &= -2\mathrm{i}\sqrt{p_{\text{L}}^2-p^2}+\frac{p\epsilon}{2(p_{\text{L}}^2-p^2)}+O(\epsilon^2).  \label{eq:kexpand4}
	\end{align}
	$ w_1 $ and $ w_2 $ are plane wave solutions propagating in the positive and negative directions, respectively.  $ w_3 $ and $ w_4 $ are exponentially divergent unphysical solutions.
\subsection{Dark soliton solution in comoving frame}
	Let us consider the dark soliton solution of stationary NLS Eq. (\ref{eq:nls2}) in the comoving frame of the soliton. In this coordinate, the soliton is static but the surrounding condensate is flowing. Let us seek the solution with the asymptotic form
	\begin{align}
		\phi(x\rightarrow \pm\infty)=\sqrt{\rho_\infty}\mathrm{e}^{\mathrm{i}(px\pm\frac{\delta}{2})}. \label{eq:nlsdsasym}
	\end{align}
	\indent It should be noted that the velocity of the soliton is not $ -p $ but $ -2p $, because the Galilean covariance of NLS equation leads to the following property:
	\begin{align}
	\begin{split}
		&\text{$ \phi(x,t) $ is a solution.} \\
		\leftrightarrow \quad & \text{$\tilde{\phi}(x,t,\alpha)=\phi(x+2\alpha t,t)\mathrm{e}^{-\mathrm{i}\alpha x}\mathrm{e}^{-\mathrm{i}\alpha^2t}$ is a solution.}
	\end{split}\label{eq:galilei}
	\end{align}
	So, if one has the solution in the form $ \phi(x,t)=\mathrm{e}^{-\mathrm{i}\mu t}\mathrm{e}^{\mathrm{i}px}f(x) $, the corresponding soliton-moving solution is given by $ \tilde{\phi}(x,t,p)=\mathrm{e}^{-\mathrm{i}(\mu-p^2)t}f(x+2pt) $. However, for brevity, we sometimes call  $ p $ ``velocity'', ignoring the difference of twice factor. \\ 
	\indent From the conservation laws of mass and momentum, one can immediately find two integration constants:
	\begin{align}
		j&=\frac{\phi^*\phi_x-\phi\phi_x^*}{2\mathrm{i}}, \\
		j_m &= |\phi_x|^2+\mu|\phi|^2-U(|\phi|^2). \label{eq:genjm}
	\end{align}
	Here $ j $ is a mass current density and $ j_m $ is a momentum current density. Let us write the density and the phase of the condensate as $ \phi=\sqrt{\rho}\mathrm{e}^{\mathrm{i}S} $. Taking account of the asymptotic form (\ref{eq:nlsdsasym}), the chemical potential $ \mu $  becomes the same as (\ref{eq:cp}), and the above constants are determined as 
	\begin{align}
		j&=\rho_\infty p, \label{eq:dsj} \\
		j_m&=2\rho_\infty p^2+\rho_\infty F(\rho_\infty)-U(\rho_\infty). \label{eq:dsjm}
	\end{align}
	The conservation laws are then rewritten as
	\begin{align}
		S_x &= \frac{j}{\rho} =\frac{\rho_\infty p}{\rho} \ \leftrightarrow \ S = p\int_0^x\frac{\rho_\infty\mathrm{d}x}{\rho}, \label{eq:phasecond} \\
		\frac{(\rho_x)^2}{4} &= -p^2(\rho_\infty-\rho)^2+\rho\left[ U(\rho)-U(\rho_\infty)-(\rho-\rho_\infty)U'(\rho_\infty) \right]. \label{eq:momconsrv}
	\end{align}
	Thus, one can at least obtain the formal solution 
	\begin{align}
		\pm 2(x-x_0) = \int\!\!\frac{\mathrm{d}\rho}{\sqrt{\text{R.H.S. of Eq. (\ref{eq:momconsrv})}}},
	\end{align}
	even though it is not easy in general to carry out this integration and obtain the solution in closed form ``$ \rho(x)=\dots $''. Henceforth we do not need this formal solution, but we assume the existence of a dark soliton solution which has no singularity and satisfies the asymptotic condition (\ref{eq:nlsdsasym}). \\
	\indent From Eq. (\ref{eq:phasecond}), The phase shift $ \delta $ in Eq. (\ref{eq:nlsdsasym}) can be written down explicitly:
	\begin{align}
		\delta = p\int_{-\infty}^\infty\!\!\mathrm{d}x\left( \frac{\rho_\infty}{\rho}-1 \right).
	\end{align}
	We also introduce the symbol for the particle number of the dark soliton for later convenience:
	\begin{align}
		N := \int_{-\infty}^\infty\!\!\mathrm{d}x\left( \rho-\rho_\infty \right)<0.
	\end{align}
\subsection{Barashenkov's criterion}
	The stability of the soliton is described by the following renormalized momentum\cite{Barashenkov}:
	\begin{align}
		P = \int_{-\infty}^\infty\mathrm{d}x\left( \frac{\tilde{\phi}^*\tilde{\phi}_x-\tilde{\phi}\tilde{\phi}^*_x}{2\mathrm{i}} \right)\left( 1-\frac{\rho_\infty}{|\tilde{\phi}|^2} \right)
	\end{align}
	Here $ \tilde{\phi}(x,t) $ is the dark soliton solution in the frame where the surrounding condensate is at rest and the soliton is moving. Writing the soliton velocity $ v $, the stability criterion for the dark solitons is expressed by $ \partial P/\partial v<0 $.\\
	\indent We can rewrite the above integral by the density profile  $ \rho(x) $:
	\begin{align}
	\begin{split}
		P &= -p\int_{-\infty}^\infty\mathrm{d}x\left( \rho_\infty-\rho \right)\left( \frac{\rho_\infty}{\rho}-1 \right) \\
		&= -p N-\rho_\infty \delta.
	\end{split}
	\end{align}
	Here remember that the soliton velocity is given by $ v=-2p $, as stated in the preceding subsection. The stability condition is rewritten as $ \partial (-P)/\partial p <0 $.
\subsection{Definition of the scattering problem}\label{sec:defofsp}
	In this subsection, we define the transmission and reflection problem of Bogoliubov phonons shown in Fig. \ref{introfigure}. Since the linearized equation does not satisfy simple particle number conservation, we must define transmission and reflection coefficients via the conservation of excitation energy. The conservation of excitation energy corresponds to the constancy of the following Wronskian:\cite{Kagan, DanshitaYokoshiKurihara}:
	\begin{align}
		W = u^*\partial_xu-u\partial_xu^*+v^*\partial_xv-v\partial_xv^*.
	\end{align}
	Let us assume that the asymptotic form of the condensate wavefunction  $ \phi(x) $ is given by Eq. (\ref{eq:nlsdsasym}). 
	In this situation, sufficiently far from the origin, the Bogoliubov equations have the plane wave (and exponentially decaying/diverging) solutions given by Eq. (\ref{eq:bogouniform}) with $ \varphi=\pm\frac{\delta}{2} $. \\
	\indent The solution of the scattering problem is defined by the one that has the following asymptotic form\cite{Kovrizhin,Kagan,DanshitaYokoshiKurihara}: 
	\begin{align}
		\begin{pmatrix}u \\ v \end{pmatrix} \rightarrow \begin{cases}  w_1\left(x,-\frac{\delta}{2}\right)+r\, w_2\left(x,-\frac{\delta}{2}\right) & (x\rightarrow-\infty) \\ t\, w_1\left(x,+\frac{\delta}{2}\right) & (x\rightarrow+\infty). \end{cases} \label{eq:uvasymptotic}
	\end{align}
	Here the exponentially decaying waves, which are $ w_4(x,-\frac{\delta}{2}) $ in $ x\rightarrow -\infty $ and $ w_3(x,+\frac{\delta}{2}) $ in  $ x\rightarrow+\infty $, can be also included. However, they are irrelevant in the definition of transmission and reflection coefficients. The calculation of  $ W $ shows that
	\begin{align}
		\frac{W(+\infty)}{2\mathrm{i}}=&|t|^2\left[ (k_{1}+p)|\bar{u}_{1}|^2+(k_{1}-p)|\bar{v}_{1}|^2 \right]\!, \\
		\begin{split}
		\frac{W(-\infty)}{2\mathrm{i}}=&(k_{1}+p)|\bar{u}_{1}|^2+(k_{1}-p)|\bar{v}_{1}|^2 \\
		&\ +|r|^2\left[ (k_{2}+p)|\bar{u}_{2}|^2+(k_{2}-p)|\bar{v}_{2}|^2 \right].
		\end{split}
	\end{align}
	Since $ W(+\infty)=W(-\infty) $, the transmission coefficient $ T $ and the reflection coefficient $ R $ are naturally defined as
	\begin{align}
		T&=|t|^2, \label{eq:tc} \\
		R&=\frac{(-k_{2}-p)|\bar{u}_{2}|^2+(-k_{2}+p)|\bar{v}_{2}|^2}{(k_{1}+p)|\bar{u}_{1}|^2+(k_{1}-p)|\bar{v}_{1}|^2}|r|^2. \label{eq:defofR}
	\end{align}
	By this definition, $ T+R=1 $ always holds. If one chooses the normalization $ \bar{u}_i $ and $ \bar{v}_i $ as Eq. (\ref{eq:ucoeff}) and (\ref{eq:vcoeff}), one can show
	\begin{align}
		\frac{(-k_{2}-p)|\bar{u}_{2}|^2+(-k_{2}+p)|\bar{v}_{2}|^2}{(k_{1}+p)|\bar{u}_{1}|^2+(k_{1}-p)|\bar{v}_{1}|^2} = 1-\frac{pp_{\text{L}}\epsilon^2}{2(p^2-p_{\text{L}}^2)^3}+O(\epsilon^4).
	\end{align}
	So, if one is only interested in the leading order, one can approximate $ R \simeq |r|^2 $. 
\section{Summary of Main Result and Numerical Verification}\label{sec:mainresult}
	In this section we present the main results of this paper and verify them by numerical study of CQNLS equation. The proof will be given in Secs.~\ref{sec:proof1}~and~\ref{sec:proof2}.
\subsection{Main result}\label{subsec:mainresult}
	In the scattering problem of linearized excitations defined in Subsec.~\ref{sec:defofsp}, the amplitude of the reflected component $ r $ in Eq. (\ref{eq:uvasymptotic}) is given by the following Pad\'e approximant-like form:
	\begin{align}
		r = \frac{- \mathrm{i}\epsilon(d+d_1P_p)+O(\epsilon^2)}{a P_p - \mathrm{i}\epsilon(b+b_1P_p)+O(\epsilon^2)}. \label{eq:mainr}
	\end{align}
	Here $ X_p:= \partial X/\partial p $ and
	\begin{align}
		a &= 4p_{\text{L}}\rho_\infty, \\
		b &= (N+pN_p)^2+(p_{\text{L}}N_p)^2, \\
		d &= (N+pN_p)^2-(p_{\text{L}}N_p)^2, \\
		b_1 &= N-\frac{p_{\text{L}}^2+p^2}{p_{\text{L}}^2-p^2}\widetilde{N}, \\
		d_1 &= N+\widetilde{N}
	\end{align}
	with
	\begin{align}
		\widetilde{N}:=p \frac{\partial N}{\partial p}-\rho_\infty\frac{\partial N}{\partial \rho_\infty}.
	\end{align}
	From (\ref{eq:mainr}), the energy dependence of the reflection coefficient $ R $ (\ref{eq:defofR}) becomes
	\begin{align}
		R = \begin{cases} \displaystyle \left( \frac{d+d_1P_p}{aP_p} \right)^2\epsilon^2+O(\epsilon^4) & (P_p\ne0) \\ \displaystyle \left( \frac{d}{b} \right)^2+O(\epsilon^2) & (P_p=0). \end{cases} \label{eq:mainrR}
	\end{align}
	When $ P_p \ne0$, the zero-energy phonon transmits perfectly: $ \lim_{\epsilon\rightarrow0}R=0 $. On the other hand, when the soliton velocity reaches a critical value, i.e.  $ P_p=0 $, this perfect transmission disappears. We note that the rational form expression (\ref{eq:mainr}) makes it possible to unify the description of low-energy behaviors in both critical and non-critical cases. If we use a simple Taylor series, the singular behavior at the critical velocity state cannot be expressed. \\
	\indent The above (\ref{eq:mainrR}) is one good result valid for any soliton velocity, even if it is far from the critical state. However, when the velocity comes close to the critical value, i.e.,  $ P_p $ comes close to zero, we can derive a more powerful scaling law as below. Let us assume that coefficients of  $ \epsilon^n \ (n\ge2) $ in (\ref{eq:mainr}) are all finite in the limit $ P_p\rightarrow0 $. and take the limit $ \epsilon\rightarrow 0 $ and $ P_p\rightarrow0 $ with a constraint $ \epsilon/P_p=\text{fix} $. We then obtain
	\begin{align}
		r \rightarrow \frac{-\mathrm{i}d(\epsilon/P_p)}{a-\mathrm{i}b(\epsilon/P_p)},
	\end{align}
	and in the same limit, the \textit{universal} form of reflection coefficient
	\begin{align}
		\lim_{\substack{\epsilon\rightarrow0,\, P_p\rightarrow0,\\ \epsilon/P_p:\text{fix}}} R = \frac{d^2(\epsilon/P_p)^2}{a^2+b^2 (\epsilon/P_p)^2} \label{eq:scaledR}
	\end{align}
	follows. Here, the values of $ a, b, $ and $ d $ in the critical state must be substituted when we use Eq. (\ref{eq:scaledR}). We remark that Eq. (\ref{eq:scaledR}) contains only $ p $-derivatives, whereas the expression before taking the scaling limit contains $ \rho_\infty $-derivatives in addition to $ p $-derivatives. \\
	\indent Let $ p_c $ be a critical velocity of the dark soliton, i.e., $ P'(p_c)=0 $. The expansion of $ P $  near $ p=p_c $ gives
	\begin{align}
		P(p) &\simeq P(p_c)+\frac{1}{2}P''(p_c)(p-p_c)^2+\dotsb, \\
		\rightarrow |P_p| &= |P'(p)| \simeq |2P''(p_c)(P(p)-P(p_c))|^{1/2}.
	\end{align}
	Therefore we obtain
	\begin{align}
		\frac{\epsilon}{|P_p|} \simeq \frac{\epsilon}{|2P''(p_c)(P(p)-P(p_c))|^{1/2}}. \label{eq:energyscale}
	\end{align}
	This is an expected scaling  behavior from the normal form of saddle-node bifurcation\cite{GuckenheimerHolmes,PhamBrachet}, if we regard the renormalized momentum $ P $ as a parameter of normal form.
\subsection{Comparison with Numerical Results in CQNLS System}\label{subsec:cqnls}
	In this subsection, we numerically verify the analytical results of the preceding subsection in the CQNLS system. We first derive the expressions for the dark soliton solution and the renormalized momentum in Subsec.~\ref{subsubsec:darksolitoncqnls}, and solve the scattering problems of linearized excitations for (i) the purely cubic case in Subsec. \ref{sec:purecubic}, (ii) the purely quintic case in Subsec. \ref{sec:purequintic}, and (iii) the case where a non-trivial  critical velocity exists in Subsec. \ref{sec:criticalcase}.
\subsubsection{Dark soliton solution and renormalized momentum}\label{subsubsec:darksolitoncqnls}
\begin{figure}[t]
		\begin{center}
		\includegraphics[scale=1.41]{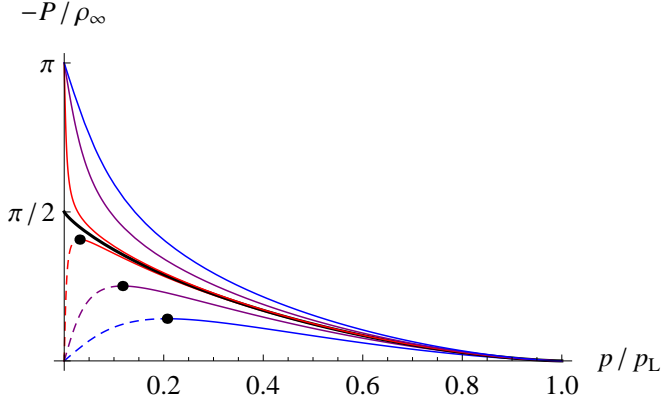}
		\caption{\label{fig:renP}(Color online) Plot of renormalized momentum $ -P/\rho_\infty=\delta+pN/\rho_\infty $ in CQNLS. Here we set $ (a_1,a_2)=(-1,1) $. The values of $ \rho_\infty $ of each curve are set, from top to bottom,  $ \rho_\infty=0.55,\, 0.52,\, 0.502,\, 0.5,\, 0.498,\, 0.48,\, \text{ and } 0.45 $, respectively. The dark solitons are stable in the regions of the solid lines, while unstable in the regions of the dashed lines. The critical points are marked by black dots. The unstable soliton appears when $ \rho_\infty<-a_1/(2a_2)=0.5 $. }
		\end{center}
	\end{figure}

	 In the CQNLS system, the nonlinear term is defined by
	\begin{align}
		U(\rho) &= a_1\rho^2+a_2\rho^3, \\
		F(\rho)&=U'(\rho) = 2a_1\rho+3a_2\rho^2,
	\end{align}
	and the NLS equation (\ref{eq:nls}) has the cubic-quintic nonlinearity:
	\begin{align}
		\mathrm{i}\partial_t\phi=-\partial_x^2\phi+2a_1|\phi|^2\phi+3a_2|\phi|^4\phi.
	\end{align}
	The stationary linearized equation, i.e., the stationary Bogoliubov equation (\ref{eq:bogos}) is given by
	\begin{align}
		\!\!\!\!\!\!&\begin{pmatrix}-\partial_x^2-\mu+4a_1|\phi|^2+9a_2|\phi|^4 & -2a_1\phi^2-6a_2|\phi|^2\phi^2 \\ 2a_1\phi^{*2}+6a_2|\phi|^2\phi^{*2} & \!\! \partial_x^2+\mu-4a_1|\phi|^2-9a_2|\phi|^4 \end{pmatrix}\begin{pmatrix}u \\ v \end{pmatrix} \nonumber \\
		&=\epsilon\begin{pmatrix}u\\v\end{pmatrix}. \label{eq:bogocqnls}
	\end{align}
	It is known that a bubble and unstable dark solitons appear when $ a_1<0 $ and $ a_2>0 $\cite{BarashenkovMakhankov,BarashenkovPanova}. This case is considered in Subsec.~\ref{sec:criticalcase}. As shown in \cite{BarashenkovPanova}, when the soliton velocity is smaller than the critical value, a small perturbation induces ``nucleation dynamics'', and the soliton cannot preserve its shape any more. So, this instability is not convective but absolute. \\
	\indent The Landau velocity (\ref{eq:defofpl}) is given by
	\begin{align}
		p_{\text{L}} = \sqrt{\rho_\infty(a_1+3a_2\rho_\infty)}.
	\end{align}
	The necessary condition  $ a_1+3a_2\rho_\infty>0 $ follows for a uniform condensate to be stable.  The dark soliton solution is given by
	\begin{align}
		\phi(x,p,\rho_\infty) =\mathrm{e}^{\mathrm{i}px}\frac{\kappa\rho_0+\mathrm{i}p(\rho_\infty-\rho_0)\tanh\kappa x}{\sqrt{\rho_0(\kappa^2-a_2(\rho_\infty-\rho_0)^2\tanh^2\kappa x)}} \label{eq:dscqnls}
	\end{align}
	with
	\begin{align}
		\kappa &=\sqrt{p_{\text{L}}^2-p^2}, \\
		\rho_0 &=\rho(x=0)= \frac{-(2a_2\rho_\infty+a_1)+\sqrt{(2a_2\rho_\infty+a_1)^2+4a_2p^2}}{2a_2}. \label{eq:rho0cqnls}
	\end{align}
	See \ref{app:cqnls} for a detailed derivation. Since $ \kappa $ and $ \rho_0 $ are the functions of $ (p,\rho_\infty) $, the dark soliton solution has two parameters $ (p,\rho_\infty) $. From (\ref{eq:rho0cqnls}), 
	\begin{align}
		\lim_{p\rightarrow0}\rho_0=\begin{cases} 0 & (2a_2\rho_\infty+a_1>0) \\ \frac{1}{2a_2}|2a_2\rho_\infty+a_1| & (2a_2\rho_\infty+a_1<0), \end{cases}
	\end{align}
	so the bubble (= a non-topological dark soliton) appears when $ \rho_\infty<-a_1/(2a_2) $. 
	A particle number of soliton $ N $ and a phase difference $ \delta $ are calculated as
	\begin{align}
	 	N &= -\frac{2}{\sqrt{a_2}}\tanh^{-1}\left[ \frac{\sqrt{a_2}(\rho_\infty-\rho_0)}{\kappa} \right],\label{eq:cqnlsN} \\
		\delta &= 2 \tan^{-1}\left[ \frac{p(\rho_\infty-\rho_0)}{\rho_0\kappa} \right]. \label{eq:cqnlsdelta}
	\end{align}
	From them we can calculate the renormalized momentum $ -P=pN+\rho_\infty \delta $. An example is shown in Fig. \ref{fig:renP}. The case where the unstable region exists is analyzed in Subsec. \ref{sec:criticalcase} in detail.
\subsubsection{Purely cubic case}\label{sec:purecubic}
	As a first example, let us consider the case $ a_2=0 $, i.e., the nonlinearity is purely cubic. As mentioned in the Introduction, the NLS equation is integrable in this case and the Bogoliubov phonons are reflectionless for any energy. Let us see that our analytical result Eq.~(\ref{eq:mainrR}) is consistent with these known facts. \\
	\indent Without loss of generality, we can set $ a_1=1 $. The dark soliton solution (\ref{eq:dscqnls}) is then reduced to 
	\begin{align}
		\phi = \mathrm{e}^{\mathrm{i}px}\left( p+\mathrm{i}\kappa\tanh \kappa x \right), \quad \kappa = \sqrt{\rho_\infty-p^2}.
	\end{align}
	The exact solution of the linearized equation, i.e., the Bogoliubov equation is given by the squared Jost solution\cite{ChenChenHuang,Kovrizhin}:
	\begin{align}
		u &= \mathrm{e}^{\mathrm{i}(k_j+p)x}\left( \mathrm{i}\kappa\tanh \kappa x+\frac{k_j}{2}+\frac{\epsilon}{2k_j} \right)^2, \\
		v &= \mathrm{e}^{\mathrm{i}(k_j-p)x}\left( \mathrm{i}\kappa\tanh \kappa x+\frac{k_j}{2}-\frac{\epsilon}{2k_j} \right)^2,
	\end{align}
	where  $ k_j $s are given by the roots of the dispersion relation Eq. (\ref{eq:uniformdisp}) with  $ F'(\rho_\infty)=2 $.\\
	\indent From the above explicit expression, it is obvious that the phonons are reflectionless. Therefore, the coefficient of $ \epsilon^2 $ in Eq.~(\ref{eq:mainrR}) must vanish. Let us check it. For the cubic case, it follows that
	\begin{align}
		N=-2\kappa,\quad \delta = 2 \tan^{-1}\frac{\kappa}{p},
	\end{align}
	by taking the limit $ a_2\rightarrow0 $  of Eqs. (\ref{eq:cqnlsN}) and (\ref{eq:cqnlsdelta}). With the use of them, we can show
	\begin{gather}
		P_p = -N-pN_p-\rho_\infty \delta_{p}=4\kappa, \\
		d=4(\kappa^2-3p^2),\quad d_1=\frac{-\kappa^2+3p^2}{\kappa}.
	\end{gather}
	Thus we obtain $ d+d_1P_p=0 $, as expected. We also note that the soliton is always stable since $ -P_p<0 $ for all velocities.\\
	\indent It is also possible to discuss the reflection properties when the quintic term is small by expanding Eqs (\ref{eq:cqnlsN}) and (\ref{eq:cqnlsdelta}) with respect to $ a_2 $, but the expression is not so simple. In this case, one can derive an approximate formula valid not only for small energy but for arbitrary energy by the method given in Refs. \cite{Muryshev,SinhaChernyKovrizhinBrand}.
\subsubsection{Purely quintic case}\label{sec:purequintic}
	Next, we treat the purely (self-defocusing) quintic case. As already mentioned, the quintic NLS equation is known to describe the dynamics of the Tonks-Girardeau gas\cite{Kolomeisky}.\\
	\indent Without loss of generality, we can set $ a_1=0 $ and $ a_2=1 $. Though we can also set $ \rho_\infty=1 $, we keep it for a moment because we need the differentiation of $ \rho_\infty $ to calculate  the reflection coefficient (\ref{eq:mainrR}).   Eqs. (\ref{eq:cqnlsN}) and (\ref{eq:cqnlsdelta}) are reduced to
	\begin{align}
		N &= -\tanh^{-1}\left[ \frac{\sqrt{3(1-y^2)}}{2} \right], \\
		\delta &= 2 \tan^{-1}\left[ \frac{1-3y^2+\sqrt{1+3y^2}}{3y\sqrt{1-y^2}} \right] \\
		\text{with}\quad y &:= \frac{p}{p_{\text{L}}} = \frac{1}{\sqrt{3}}\frac{p}{\rho_\infty}.
	\end{align}
	From them one can plot the renormalized momentum $ -P/\rho_\infty=\sqrt{3}yN+\delta $ and can show that $ -P_p<0 $ always holds, i.e., the soliton is always stable. One can also obtain the coefficient of $ \epsilon^2 $ in Eq.~(\ref{eq:mainrR}) as follows:
	\begin{align}
		-\frac{d+d_1P_p}{aP_p} &= \frac{1}{4\sqrt{3}\rho_\infty^2}\frac{\gamma\tanh^{-1}\gamma}{\gamma+2(1-\gamma^2)\tanh^{-1}\gamma}, \label{eq:quinticcoeff} \\
		 \gamma&:=\frac{\sqrt{3(1-y^2)}}{2}.
	\end{align}
	\begin{figure}[t]
		\begin{center}
		\includegraphics[scale=0.87]{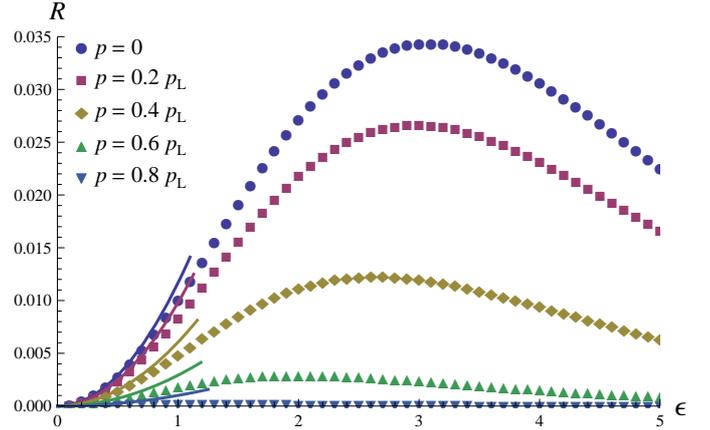}
		\caption{\label{fig:quinticRC}(Color online) Energy-dependence of reflection coefficient $ R $ of linearized excitations for various soliton velocities in the purely quintic system. Here we set $ (a_1,a_2)=(0,1) $ and $ \rho_\infty=1 $. The Landau velocity is given by $ p_{\text{L}}=\sqrt{3}\rho_\infty=\sqrt{3} $. Parabolic curves represent the theoretical approximate expression (\ref{eq:mainrR}) with (\ref{eq:quinticcoeff}).}
		\end{center}
	\end{figure}
	The energy-dependence of the reflection coefficient $ R $ of linearized excitations is obtained by solving the Bogoliubov equation (\ref{eq:bogocqnls}) numerically, and the results are shown in Fig.~\ref{fig:quinticRC}.
	We can verify that the expression (\ref{eq:mainrR}) with (\ref{eq:quinticcoeff}) is valid for low-energy region. From this figure we can also see that the soliton with zero velocity is the strongest scatterer. It is intuitively clear since the shape of the soliton becomes shallower and wider if the velocity of the soliton increases. However, this intuitive understanding is not always correct, as the integrable cubic case in Subsec.~\ref{sec:purecubic} and the instability-induced anomaly in Subsec.~\ref{sec:criticalcase} illustrate.
\subsubsection{The case with $ a_1<0 $ and $ a_2>0 $}\label{sec:criticalcase}
\begin{figure}[t]
		\begin{center}
		\includegraphics[scale=0.87]{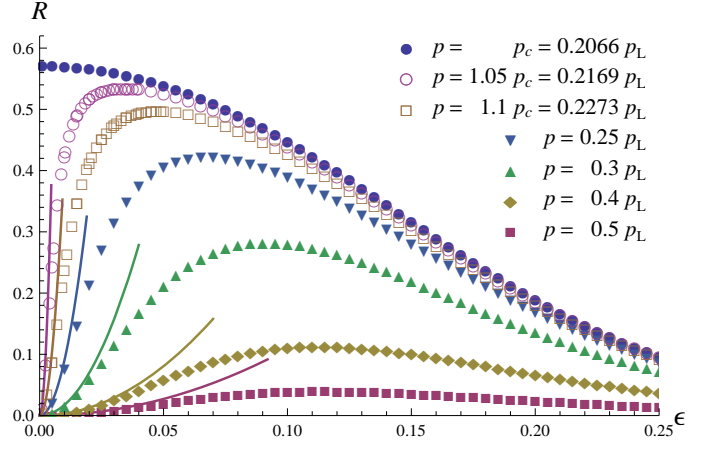}
		\caption{\label{fig:RC}(Color online) Energy-dependence of reflection coefficient $ R $ of linearized excitations for various soliton velocities. Here we set $ (a_1,a_2)=(-1,1) $ and $ \rho_\infty=0.45 $. The critical velocity of the dark soliton is given by $ p_c = (0.206597\dots)\times p_{\text{L}}$. (See the lowest curve of Fig. \ref{fig:renP}.) A reflection coefficient of zero-energy phonon at the critical state is given by $ (d/b)^2\simeq 0.5718 $. Parabolic curves represent the theoretical approximate expression (\ref{eq:mainrR}).}
		\end{center}
	\end{figure}

\begin{figure}[t]
		\begin{center}
		\includegraphics[scale=0.87]{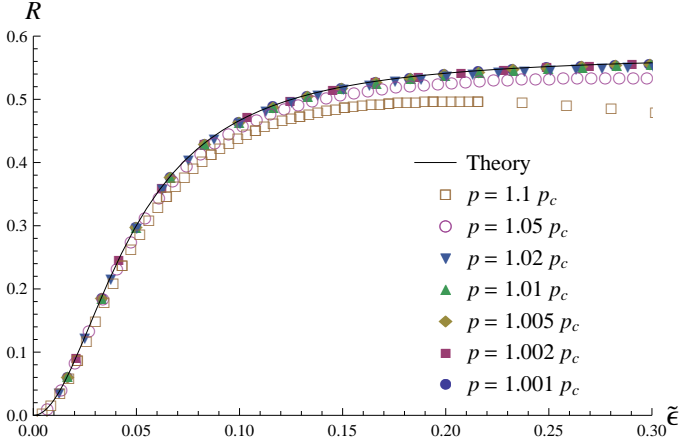}
		\caption{\label{fig:scaledR}(Color online) Scaling behavior of reflection coefficient $ R $. Here we set  $ (a_1,a_2)=(-1,1) $ and $ \rho_\infty=0.45 $. $ \tilde{\epsilon}:=\epsilon/\sqrt{2P''(p_c)(P(p)-P(p_c))} $ is a scaled excitation energy. ``Theory'' represents the universal form of reflection coefficient (\ref{eq:scaledR}).}
		\end{center}
	\end{figure}

	Finally, we consider the case with $ a_1<0 $ and $ a_2>0 $, which is most interesting from the viewpoint of critical phenomena, since the soliton can become unstable and the reflection coefficient can show the singular and scaling behavior. \\
	\indent If both $ a_1 $ and $ a_2 $ are nonzero, we can set $ |a_1|=|a_2|=1 $ without loss of generality by the following scale transformation:
	\begin{gather}
		\bar{x}=\frac{x}{\xi},\ \bar{t}=\frac{t}{\xi^2},\  \bar{\phi}(\bar{x},\bar{t})=\frac{1}{\eta}\phi(x,t), \\
		\xi=\frac{\sqrt{|a_2|}}{|a_1|},\ \eta=\sqrt{\frac{|a_1|}{|a_2|}}.
	\end{gather}
	So we performed numerical calculations by setting $ (a_1,a_2)=(-1,1)$.  Note that  $ \rho_\infty $ cannot be normalized to be unity if we choose  $ \xi $ and $ \eta $ as the above.  Another choice of $ \eta $ is possible to normalize $ \rho_\infty=1 $, but in this case either $ a_1 $ or $ a_2 $ cannot be normalized.\\
	\indent Using the dark soliton solution (\ref{eq:dscqnls}), we numerically solved the stationary Bogoliubov equation (\ref{eq:bogocqnls}), and constructed the solution with the asymptotic form  (\ref{eq:uvasymptotic}). Figure \ref{fig:RC} shows the reflection coefficient with $ \rho_\infty=0.45 $ for various soliton velocities. We can observe that the zero-energy phonon transmits perfectly, unless the soliton velocity is equal to the critical one. When the soliton velocity comes close to the critical value, the slope of the reflection coefficient becomes very steep, and at the critical state, the perfect transmission eventually vanishes. The approximate expression (\ref{eq:mainrR}) is good for sufficiently low excitation energy.  Figure \ref{fig:scaledR} shows the scaling behavior of reflection coefficient $ R $. Here, based on Eq. (\ref{eq:energyscale}), the horizontal axis is chosen to be the scaled energy $ \tilde{\epsilon}=\epsilon/\sqrt{2P''(p_c)(P(p)-P(p_c))} $. We can see that if the soliton velocity is close to the critical one, numerically calculated points are well fitted to the universal curve (\ref{eq:scaledR}). Thus, the theoretical results are well confirmed in this example.
\section{Proof -- Step 1: Exact Zero-Energy Solutions}\label{sec:proof1}
	In this and the next section, we prove the main result. This section is particularly devoted to the construction of exact zero-energy solutions. As an important tool, parameter derivatives are introduced.
\subsection{Parameter derivative}\label{subsec:paradera}
	As seen in the asymptotic form (\ref{eq:nlsdsasym}) or in the example of the CQNLS system in Subsec.~\ref{subsec:cqnls}, the dark soliton solution has two parameters, i.e.,  $ (p,\rho_\infty) $.  So we can consider two kinds of parameter derivatives:\footnote{Here we do not consider the derivative with respect to the parameters included in the definition of the nonlinear term, e.g.,  $ a_1 $ and $ a_2 $ in the CQNLS equation, because these parameter-derivatives  do not yield a solution of the linearized equation.}  $ \partial_p\phi $ and $ \partial_{\rho_\infty}\phi $. We can use arbitrary coordinates to ``label'' the two-dimensional parameter space $ (\alpha,\beta)=(\alpha(p,\rho_\infty), \beta(p,\rho_\infty)) $, unless the Jacobian of coordinate transformation is singular. Obviously, the final result must not depend on the choice of coordinates. In order to make the story general, we always use these general parameter derivatives, and henceforth, we write the parameter derivative simply by the subscript, i.e., $ \phi_\alpha := \partial_\alpha\phi $ and $ \phi_\beta:=\partial_\beta\phi $. We also introduce the following symbol:
	\begin{align}
		[A,B]_{\alpha\beta}:=A_\alpha B_\beta-A_\beta B_\alpha.
	\end{align}
	Note that the ratio
	\begin{align}
		\frac{[A,B]_{\alpha\beta}}{[C,D]_{\alpha\beta}} \label{eq:invratio}
	\end{align}
	has a coordinate-free meaning, in other words, it is invariant under coordinate transformations of parameter space. We often construct coordinate-free solutions in such a ratio form. \\ 
	\indent An immediate application of parameter derivatives is that one can obtain a particular solution of zero-energy Bogoliubov equation (\ref{eq:bogos}). By differentiation of the stationary NLS eq. (\ref{eq:nls2}), one obtains
	\begin{align}
		\mathcal{L}_\mu \begin{pmatrix} \phi_\alpha \\ -\phi^*_\alpha \end{pmatrix} = \mu_\alpha \begin{pmatrix} \phi \\ \phi^* \end{pmatrix}. \label{eq:1storder}
	\end{align}
	The same expression follows by replacement $ \alpha\rightarrow\beta $. Thus, taking the difference of double parameter derivatives, one can obtain the following zero-energy solution:
	\begin{align}
		\mathcal{L}_\mu \begin{pmatrix} [\mu,\phi]_{\alpha\beta} \\ -[\mu,\phi^*]_{\alpha\beta} \end{pmatrix} = 0.
	\end{align}
	It must be emphasized that this solution exists even when a localized potential barrier is added, in other words, when the fundamental equation loses a translational symmetry. What we only need is two kinds of parameter derivatives. So, this solution is not a symmetry-originated zero-mode. (For a symmetry consideration, see \ref{app:symmetry}.) \\ 
	\indent Some technical (but crucially important) identities are derived in \ref{app:idnty}. Equation (\ref{eq:1storder}) will be used again in the process of energy expansions.
\subsection{Density fluctuation and phase fluctuation}\label{subsec:fg}
	\indent Here we introduce notations for the linearized density fluctuations and phase fluctuations, and rewrite the Bogoliubov equation with respect to these variables. They are convenient for both calculations and physical interpretations. Through the symbols $ (u,v)=(\delta\phi,-\delta\phi^*) $, the density and phase fluctuations are expressed as
	\begin{align}
		\delta \rho &= \delta(\phi\phi^*)=\delta\phi\phi^*+\phi\delta\phi^*=u\phi^*-v\phi \\
		\delta S &= \delta\left( \frac{1}{2\mathrm{i}}\log\frac{\phi}{\phi^*} \right)=\frac{1}{2\mathrm{i}}\left( \frac{\delta\phi}{\phi}-\frac{\delta\phi^*}{\phi^*} \right)=\frac{1}{2\mathrm{i}}\left( \frac{u}{\phi}+\frac{v}{\phi^*} \right)
	\end{align}
	Therefore, if one defines  $ f $ and $ g $  as
	\begin{align}
		\begin{cases} \displaystyle f=\frac{1}{2\mathrm{i}}\left( \frac{u}{\phi}+\frac{v}{\phi^*} \right) \\ g=u\phi^*-v\phi \end{cases} \quad\leftrightarrow\quad \begin{cases}\displaystyle  u=\phi\left( \mathrm{i}f+\frac{g}{2\rho} \right) \\ \displaystyle v=\phi^*\left( \mathrm{i}f-\frac{g}{2\rho} \right), \end{cases}
	\end{align}
	then  $ f $  has the meaning of the phase fluctuation, and  $ g $ the density fluctuation. The stationary Bogoliubov equation (\ref{eq:bogos}) are rewritten as follows:
	\begin{align}
		\left( \rho f'+\frac{jg}{\rho} \right)'&=\frac{\mathrm{i}\epsilon}{2}g, \label{eq:bogofg1}\\
		\left( g'-\frac{\rho'}{\rho}g \right)'-2\rho F'(\rho)g-4jf'&=-2\mathrm{i}\epsilon\rho f \label{eq:bogofg2}.
	\end{align}
	For the equation with $ \epsilon=0 $, we obtain the following integration constant:
	\begin{align}
		\rho f'+\frac{jg}{\rho}=\text{const.} \label{eq:zeroBogoconst}
	\end{align}
\subsection{Exact zero-energy solutions}\label{subsec:zerosol}
	Let us derive all four linearly independent solutions for the zero-energy Bogoliubov equation (\ref{eq:bogos}), or equivalently, Eqs. (\ref{eq:bogofg1}) and (\ref{eq:bogofg2}) with  $ \epsilon=0 $. From global phase symmetry and translational symmetry, we can immediately find two zero-modes:  $ (u,v)=(\mathrm{i}\phi,\mathrm{i}\phi^*) $ and $ (\phi^{}_x,-\phi^*_x) $. In addition to them, we already have the third solution in Subsec. \ref{subsec:paradera}. It is well known that if we have $ n-1 $ linearly independent solutions for an $ n $-th order linear differential equation, the last one can be obtained by reduction of order. So, the fourth solution also can be found. See \ref{app:roo} for a detailed calculation. \\ 
	\indent Thus, all four zero-energy solutions are given by
	\begin{align}
		\begin{pmatrix}f_1 \\ g_1 \end{pmatrix}&=\begin{pmatrix}1 \\ 0 \end{pmatrix}, \\
		\begin{pmatrix}f_2 \\ g_2 \end{pmatrix}&=\frac{1}{[\mu,j]_{\alpha\beta}}\begin{pmatrix}[\mu,S]_{\alpha\beta} \\ [\mu,\rho]_{\alpha\beta} \end{pmatrix},\\
		\begin{pmatrix}f_3 \\ g_3 \end{pmatrix}&=\begin{pmatrix}S_x-p \\ \rho_x \end{pmatrix}=\begin{pmatrix}p(\rho_\infty-\rho)/\rho \\ \rho_x \end{pmatrix}, \label{eq:f3g3} \\
		\begin{pmatrix}f_4 \\ g_4 \end{pmatrix}&= -\frac{\rho g_2}{\rho_\infty-\rho}\begin{pmatrix}0 \\ 1\end{pmatrix}+\left[\rho_\infty \int_0^x\!\!\frac{ g_2\mathrm{d}x}{(\rho_\infty-\rho)^2} \right]\begin{pmatrix} f_3 \\ g_3 \end{pmatrix}. \label{eq:zerosol4}
	\end{align}
	For reference, we also write down the same solutions by $ (u,v) $ notation:
	\begin{align}
		\begin{pmatrix}u_1 \\ v_1 \end{pmatrix}&=\begin{pmatrix}\mathrm{i}\phi \\ \mathrm{i}\phi^* \end{pmatrix}, \\
		\begin{pmatrix}u_2 \\ v_2 \end{pmatrix}&=\frac{1}{[\mu,j]_{\alpha\beta}}\begin{pmatrix} [\mu,\phi]_{\alpha\beta} \\ -[\mu,\phi^*]_{\alpha\beta} \end{pmatrix},\\
		\begin{pmatrix}u_3 \\ v_3 \end{pmatrix}&=\begin{pmatrix}\phi_x-\mathrm{i}p\phi \\ -\phi_x^*-\mathrm{i}p\phi^* \end{pmatrix}, \label{eq:u3v3}\\
		\begin{pmatrix}u_4 \\ v_4 \end{pmatrix}&= \frac{g_2}{2(\rho_\infty-\rho)}\begin{pmatrix}-\phi \\ \phi^*\end{pmatrix}+\left[\rho_\infty \int_0^x\!\!\frac{ g_2\mathrm{d}x}{(\rho_\infty-\rho)^2} \right]\begin{pmatrix} u_3 \\ v_3 \end{pmatrix}
	\end{align}
	Here $ g_2 = [\mu,\rho]_{\alpha\beta}/[\mu,j]_{\alpha\beta} = \mathrm{i}(u_1v_2-u_2v_1) $. \\
	\indent $ (f_4,g_4) $ is chosen so that the integration constant (\ref{eq:zeroBogoconst}) vanishes. So, only $ (f_2,g_2) $ contributes to this constant:
	\begin{align}
		\rho f_2'+\frac{jg_2}{\rho}&=1, \\
		\rho f_i'+\frac{jg_i}{\rho}&=0, \quad (i=\text{1, 3, and 4.}). 
	\end{align}
	\indent If we assume that the density $ \rho(x) $ is an even function, the parities and asymptotic behaviors of the four solutions are summarized as follows:
	\begin{align}
		(f_1,g_1)\, &: \ (\text{even},\text{odd}), \quad \text{bounded.} \label{eq:zerosolparity1} \\
		(f_2,g_2)\, &: \ (\text{odd},\text{even}), \quad \text{linearly divergent.} \\
		(f_3,g_3)\, &: \ (\text{even},\text{odd}), \quad \text{exponentially decreasing.} \\
		(f_4,g_4)\, &: \ (\text{odd},\text{even}), \quad \text{exponentially increasing.} \label{eq:zerosolparity4}
	\end{align}
	Keeping in mind the above (\ref{eq:zerosolparity1})--(\ref{eq:zerosolparity4}) will help to understand and imagine the construction of the solution in the next section. More detailed forms of asymptotes are given in \ref{sec:asymofzero}. \\
	\indent Deriving all solutions including divergent ones might seem to be only of academic interest, but they will become necessary for the exact energy expansion, which will be performed in the next section. \\
	\indent We note that four solutions are also obtained without using the parameter derivatives as shown in \ref{app:zerosol}. However, those expressions are almost useless for solving a scattering problem. 
\section{Proof -- Step 2: Low-Energy Expansion}\label{sec:proof2}
	In this section, with the use of zero-energy solutions derived in the preceding section, we construct finite-energy solutions up to second order by an exact energy expansion method. Using them, we solve the scattering problem of linearized excitations, and prove the main result.
\subsection{Exact low-energy expansion}
	We construct the finite energy solution by energy expansion:
	\begin{align}
		\begin{pmatrix}u \\ v \end{pmatrix} = \sum_{n=0}^\infty \epsilon^n \begin{pmatrix} u^{(n)} \\ v^{(n)} \end{pmatrix}, \label{eq:expansion}
	\end{align}
	or equivalently,  $ (f,g) = \sum_{n=0}^\infty \epsilon^n (f^{(n)},g^{(n)}) $. The expansion only using asymptotic forms was used in Ref.~\cite{KatoNishiwakiFujita}, and the completely exact expansion was first used in Ref.~\cite{TakahashiKato} to discuss tunneling behaviors of Bogoliubov phonons in the presence of potential walls. In order to prove the singular behavior at the critical point, the exact expansion is essentially necessary \cite{TakahashiKato,TakahashiKatoConf}.\\
	\indent The substitution of (\ref{eq:expansion}) into Bogoliubov equation (\ref{eq:bogos}) yields
	\begin{align}
		\mathcal{L}_\mu\begin{pmatrix}u^{(n)} \\ v^{(n)} \end{pmatrix} = \begin{pmatrix}u^{(n-1)} \\ v^{(n-1)} \end{pmatrix}, \quad (n=1,2,\dots), \label{eq:expansion2}
	\end{align}
	or in $ (f,g) $ expression,
	\begin{align}
		\left( \rho \left(f^{(n)}\right)'+\frac{jg^{(n)}}{\rho} \right)'&=\frac{\mathrm{i}}{2}g^{(n-1)}, \label{eq:recurf} \\
		\left( \left(g^{(n)}\right)'-\frac{\rho'}{\rho}g^{(n)} \right)'-2\rho F'(\rho)g^{(n)}-4j\left(f^{(n)}\right)' &=-2\mathrm{i}\rho f^{(n-1)}. \label{eq:recurg}
	\end{align}
	It is equivalent to the zero-energy Bogoliubov equation with an inhomogeneous term  $ (u^{(n-1)},v^{(n-1)}) $. Since we already know all four solutions for the homogeneous equation, we can solve it by variation of parameters. Since we want the solution of the form (\ref{eq:uvasymptotic}), in the process of energy expansions, we must cancel out the exponentially divergent terms, and must take up the power-law divergent terms, because $ \mathrm{e}^{\mathrm{i}k_ix}=1+\mathrm{i}k_ix-k_i^2x^2/2+\dotsb $ and  $ k_1,k_2 \propto \epsilon $ for low-energy. (See eqs. (\ref{eq:kexpand1}) and (\ref{eq:kexpand2}).)
\subsection{First order solutions}
	The first order solutions are, in fact, directly calculated without the use of variation of parameters. When we choose $ (u^{(0)},v^{(0)})=(u_1,v_1)=(\mathrm{i}\phi,\mathrm{i}\phi^*) $, Eq. (\ref{eq:1storder}) gives the particular solution for $ (u^{(1)},v^{(1)}) $, that is, 
		\begin{align}
		\begin{pmatrix}\tilde{u}_{\text{A}}^{(\text{finite }\epsilon)} \\ \tilde{v}_{\text{A}}^{(\text{finite }\epsilon)} \end{pmatrix} &:= \begin{pmatrix} \mathrm{i}\phi \\ \mathrm{i}\phi^* \end{pmatrix} + \frac{\mathrm{i}\epsilon}{\mu_\alpha}\begin{pmatrix} \phi_\alpha \\ -\phi_\alpha^* \end{pmatrix}+O(\epsilon^2)
		\intertext{and}
		\begin{pmatrix}\tilde{u}_{\text{B}}^{(\text{finite }\epsilon)} \\ \tilde{v}_{\text{B}}^{(\text{finite }\epsilon)} \end{pmatrix} &:=  \begin{pmatrix} \mathrm{i}\phi \\ \mathrm{i}\phi^* \end{pmatrix} + \frac{\mathrm{i}\epsilon}{\mu_\beta}\begin{pmatrix} \phi_\beta \\ -\phi_\beta^* \end{pmatrix}+O(\epsilon^2)
	\end{align}
	give two kinds of first order solutions of the finite energy solution. From Galilean symmetry of the NLS equation, we can also obtain the first order solution when we set $ (u^{(0)},v^{(0)})=(\phi_x,-\phi^*_x) $. (See \ref{app:symmetry}) It is given by
	\begin{align}
		\begin{pmatrix}\tilde{u}_3^{(\text{finite }\epsilon)} \\ \tilde{v}_3^{(\text{finite }\epsilon)} \end{pmatrix} &:= \begin{pmatrix} \phi_x \\ -\phi^*_x \end{pmatrix} + \frac{\mathrm{i}\epsilon x}{2}\begin{pmatrix} \mathrm{i}\phi \\ \mathrm{i}\phi^* \end{pmatrix}+O(\epsilon^2). \label{eq:1stu3v3}
	\end{align}
	We also write down them in $ (f,g) $ notation:
	\begin{align}
		\begin{pmatrix}\tilde{f}_{\text{A}}^{(\text{finite }\epsilon)} \\ \tilde{g}_{\text{A}}^{(\text{finite }\epsilon)} \end{pmatrix} &:= \begin{pmatrix} 1 \\ 0 \end{pmatrix} + \frac{\mathrm{i}\epsilon}{\mu_\alpha}\begin{pmatrix} S_\alpha \\ \rho_\alpha \end{pmatrix}+O(\epsilon^2), \label{eq:firsta}\\
		\begin{pmatrix}\tilde{f}_{\text{B}}^{(\text{finite }\epsilon)} \\ \tilde{g}_{\text{B}}^{(\text{finite }\epsilon)} \end{pmatrix} &:= \begin{pmatrix} 1 \\ 0 \end{pmatrix} + \frac{\mathrm{i}\epsilon}{\mu_\beta}\begin{pmatrix} S_\beta \\ \rho_\beta \end{pmatrix}+O(\epsilon^2), \label{eq:firstb}\\
		\begin{pmatrix}\tilde{f}_3^{(\text{finite }\epsilon)} \\ \tilde{g}_3^{(\text{finite }\epsilon)} \end{pmatrix} &:= \begin{pmatrix} S_x \\ \rho_x \end{pmatrix} + \frac{\mathrm{i}\epsilon x}{2}\begin{pmatrix} 1 \\ 0 \end{pmatrix}+O(\epsilon^2). \label{eq:first3}
	\end{align}

\subsection{Second order calculation -- Identification of bounded solutions}\label{subsec:boundedsols}
	Let us calculate the second order. Before beginning, we explain the outline of the calculation. It is important to notice that only two of four solutions are the bounded solutions, i.e., propagating waves for finite positive energy $ \epsilon $, as seen in Subsec.~\ref{subsec:uniform}. This means, in the energy expansion, that the two of four solutions are power-law divergent and the remaining two must be exponentially divergent. On the other hand, however,  we have \textit{three} linearly divergent solutions (\ref{eq:firsta})--(\ref{eq:first3}) in the previous subsection. So, one of the three must be a dummy. In fact, the solutions calculated from the variation of parameters shows that all three solutions diverge exponentially; The parity of the second-order exponential divergent term becomes $ (f^{(2)},g^{(2)})=(\text{even},\text{odd}) $, so it cannot be canceled out by adding the homogeneous solution $ (f_4,g_4) $. In order to kill out this exponential divergence, we must choose the special linear combinations of the three. Through this process, we have \textit{two} power-law divergent solutions, as expected. \\ 
	\indent Now, let us construct the solution of Eqs. (\ref{eq:recurf}) and (\ref{eq:recurg}) with $ n=2 $ by the above-mentioned method. In accordance with the method of variation of parameters, let the particular solution be $ (f^{(2)},g^{(2)})=\sum_{i=1,2,3,4}c_i(f_i,g_i) $, and let $ c_i $s have the $ x $-dependence. After a little calculation, the equations for $ c_i $s are summarized as follows:
	\begin{align}
		2\mathrm{i}\begin{pmatrix}c_1' \\ c_2' \end{pmatrix} &= \begin{pmatrix} f_2 & -g_2 \\ -f_1 & g_1 \end{pmatrix}\begin{pmatrix} g^{(1)} \\ f^{(1)} \end{pmatrix}, \label{eq:vopc12}\\  
		2\mathrm{i}p\begin{pmatrix}c_3' \\ c_4' \end{pmatrix} &= \begin{pmatrix} f_4 & -g_4 \\ -f_3 & g_3 \end{pmatrix}\begin{pmatrix} g^{(1)} \\ f^{(1)} \end{pmatrix}. \label{eq:vopc34}
	\end{align}
	Let us assume that $ f^{(1)} $ is a linearly divergent odd function and $ g^{(1)} $ is a bounded even function. From the asymptotic behaviors (\ref{eq:zerosolparity1})--(\ref{eq:zerosolparity4}) and \ref{sec:asymofzero},  we can see that $ c_1(f_1,g_1),\,c_2(f_2,g_2) $, and  $ c_3(f_3,g_3) $ are always power-law divergent functions.  $ c_4 $ is an asymptotically constant function, so the product $ c_4(f_4,g_4) $ diverges exponentially in general. Exceptionally, if $ c_4 $ decays exponentially, $ c_4(f_4,g_4) $ becomes a power-law divergent function. Therefore, in order for the particular solution to be a power-law divergent function, $ \lim_{|x|\rightarrow\infty}c_4=0 $, or equivalently,
	\begin{align}
		\int_{-\infty}^{\infty}c_4'\mathrm{d}x=0 \quad\leftrightarrow\quad  \int_{-\infty}^{\infty} \left(f_3 g^{(1)}-g_3 f^{(1)}\right)\mathrm{d}x=0 \label{eq:nondiverge}
	\end{align}
	must hold.\\
	\indent  Let us construct the solution that satisfies the above condition (\ref{eq:nondiverge}). Consider the following linear combination:
	\begin{align}
	\begin{split}
		\begin{pmatrix}f_{\text{A}}^{(\text{finite }\epsilon)} \\ g_{\text{A}}^{(\text{finite }\epsilon)} \end{pmatrix} &:= \begin{pmatrix}\tilde{f}_{\text{A}}^{(\text{finite }\epsilon)} \\ \tilde{g}_{\text{A}}^{(\text{finite }\epsilon)} \end{pmatrix}+2\xi_{\text{A}}\begin{pmatrix}\tilde{f}_3^{(\text{finite }\epsilon)} \\ \tilde{g}_3^{(\text{finite }\epsilon)} \end{pmatrix} \\
		&=\begin{pmatrix} 1+2\xi_{\text{A}} S_x \\ 2\xi_{\text{A}} \rho_x \end{pmatrix} + \frac{\mathrm{i}\epsilon}{\mu_\alpha}\begin{pmatrix} S_\alpha+\mu_\alpha\xi_{\text{A}}x \\ \rho_\alpha \end{pmatrix}+O(\epsilon^2).
	\end{split}
	\end{align}
	$ \xi_{\text{A}} $ which satisfies Eq. (\ref{eq:nondiverge}) is determined as
	\begin{align}
		&\int_{-\infty}^\infty\left[ f_3\rho_\alpha-g_3(S_\alpha+\mu_\alpha\xi_{\text{A}}x) \right]\mathrm{d}x=0 \\
		\leftrightarrow \ & \xi_{\text{A}} = \frac{\rho_\infty\delta_\alpha+p N_\alpha}{N \mu_\alpha}.
	\end{align}
	The same calculation can be performed with respect to the other parameter $ \beta $, so  $ (f_{\text{B}}^{(\text{finite }\epsilon)},g_{\text{B}}^{(\text{finite }\epsilon)}) $ and $ \xi_{\text{B}} $ are defined in the same manner.\\
	\indent Thus, we have two non-exponentially-divergent solutions $ (f_{\text{A}}^{(\text{finite }\epsilon)},g_{\text{A}}^{(\text{finite }\epsilon)}) $ and $ (f_{\text{B}}^{(\text{finite }\epsilon)},g_{\text{B}}^{(\text{finite }\epsilon)}) $. Furthermore, we can get coordinate-free solutions by making the linear combinations of these two. They are defined as follows:
	\begin{align}
		\begin{pmatrix} f_1^{(\text{finite }\epsilon)} \\ g_1^{(\text{finite }\epsilon)} \end{pmatrix} &:= \frac{1}{\xi_{\text{A}}-\xi_{\text{B}}}\left[ \xi_{\text{A}}\begin{pmatrix}f_{\text{B}}^{(\text{finite }\epsilon)} \\ g_{\text{B}}^{(\text{finite }\epsilon)} \end{pmatrix}-\xi_{\text{B}}\begin{pmatrix}f_{\text{A}}^{(\text{finite }\epsilon)} \\ g_{\text{A}}^{(\text{finite }\epsilon)} \end{pmatrix} \right], \\
		\begin{pmatrix} f_3^{(\text{finite }\epsilon)} \\ g_3^{(\text{finite }\epsilon)} \end{pmatrix} &:= \frac{1}{2(\xi_{\text{A}}-\xi_{\text{B}})}\left[ \begin{pmatrix}f_{\text{A}}^{(\text{finite }\epsilon)} \\ g_{\text{A}}^{(\text{finite }\epsilon)} \end{pmatrix}-\begin{pmatrix}f_{\text{B}}^{(\text{finite }\epsilon)} \\ g_{\text{B}}^{(\text{finite }\epsilon)} \end{pmatrix} \right]-p\begin{pmatrix} f_1^{(\text{finite }\epsilon)} \\ g_1^{(\text{finite }\epsilon)} \end{pmatrix}.
	\end{align}
	If we write
	\begin{align}
		\begin{pmatrix} f_i^{(\text{finite }\epsilon)} \\ g_i^{(\text{finite }\epsilon)} \end{pmatrix} = \sum_{n=0}^\infty \epsilon^n \begin{pmatrix} f_i^{(n)} \\ g_i^{(n)} \end{pmatrix} \quad (i=\text{1 or 3}),
	\end{align}
	then the zeroth and first order terms are given by
	\begin{align}
		\begin{pmatrix} f_1^{(0)} \\ g_1^{(0)} \end{pmatrix} &= \begin{pmatrix} f_1 \\ g_1 \end{pmatrix} = \begin{pmatrix} 1 \\ 0 \end{pmatrix}, \\
		\begin{pmatrix} f_1^{(1)} \\ g_1^{(1)} \end{pmatrix} &= \frac{\mathrm{i}}{\rho_\infty[\delta,\mu]_{\alpha\beta}+p[N,\mu]_{\alpha\beta}}\begin{pmatrix} \rho_\infty[\delta,S]_{\alpha\beta}+p[N,S]_{\alpha\beta} \\ \rho_\infty[\delta,\rho]_{\alpha\beta}+p[N,\rho]_{\alpha\beta} \end{pmatrix}, \label{eq:f11g11}\\
		\begin{pmatrix} f_3^{(0)} \\ g_3^{(0)} \end{pmatrix} &= \begin{pmatrix} f_3 \\ g_3 \end{pmatrix} = \begin{pmatrix} S_x-p \\ \rho_x \end{pmatrix}, \\
		\begin{pmatrix} f_3^{(1)} \\ g_3^{(1)} \end{pmatrix} &= \frac{\mathrm{i}x}{2}\begin{pmatrix}1 \\ 0 \end{pmatrix}-\frac{\mathrm{i}N[\mu,j]_{\alpha\beta}}{2(\rho_\infty[\delta,\mu]_{\alpha\beta}+p[N,\mu]_{\alpha\beta})}\begin{pmatrix} f_2 \\ g_2 \end{pmatrix}-p\begin{pmatrix} f_1^{(1)} \\ g_1^{(1)} \end{pmatrix}\label{eq:f31g31},
	\end{align}
	and the second order terms are, from Eqs. (\ref{eq:vopc12}) and (\ref{eq:vopc34}),  given by
	\begin{align}
	\begin{split}
		\begin{pmatrix} f_i^{(2)} \\ g_i^{(2)} \end{pmatrix} &= \left[ \frac{1}{2\mathrm{i}}\int_0^x\left(f_2g_i^{(1)}-g_2f_i^{(1)}\right)\mathrm{d}x \right]\begin{pmatrix}f_1 \\ g_1 \end{pmatrix}\\
		&\quad+\left[ \frac{-1}{2\mathrm{i}}\int_0^xg_i^{(1)}\mathrm{d}x \right]\begin{pmatrix}f_2 \\ g_2 \end{pmatrix} \\
		&\qquad+\left[ \frac{1}{2\mathrm{i}p}\int_0^x\left(f_4g_i^{(1)}-g_4f_i^{(1)}\right)\mathrm{d}x \right]\begin{pmatrix}f_3 \\ g_3 \end{pmatrix} \\
		&\qquad\quad+\left[ \frac{-1}{2\mathrm{i}p}\int_0^x\left(f_3g_i^{(1)}-g_3f_i^{(1)}\right)\mathrm{d}x \right]\begin{pmatrix}f_4 \\ g_4 \end{pmatrix}
	\end{split}\label{eq:2ndbound}
	\end{align}
	with $ i=\text{1 or 3} $.\\ 
	\indent \textit{Every bounded solution for finite positive energy $ \epsilon $ must be constructed as a linear combination of  $ (f_1^{(\mathrm{finite \ }\epsilon)}, g_1^{(\mathrm{finite \ }\epsilon)}) $ and $ (f_3^{(\mathrm{finite \ }\epsilon)}, g_3^{(\mathrm{finite \ }\epsilon)}) $}, so must the solution of the scattering problem (\ref{eq:uvasymptotic}). Our remaining work is to calculate their asymptotic behavior.
\subsection{Second order calculation -- Asymptotics}\label{subsec:2ndasym}
	The evaluation of asymptotic form of the second order term (\ref{eq:2ndbound}) is tedious but straightforward. All calculations can be carried out by using the expressions in \ref{sec:asymofzero}. For brevity, we introduce the following symbols for the asymptotic forms of $ f^{(1)}_i $ and $ \int_0^x g^{(1)}_i\mathrm{d}x  $  $  (i=1\text{ or }3.)$:
	\begin{align}
		f^{(1)}_i &\rightarrow \mathrm{i}(l_{i1}x+l_{i0}\operatorname{sgn}x), \label{eq:f1asym} \\
		\int_0^xg^{(1)}_i\mathrm{d}x &\rightarrow \mathrm{i}(m_{i1}x+m_{i0}\operatorname{sgn}x). \label{eq:g1asym}
	\end{align}
	Explicit expressions for $ l_{ij}s $ and $ m_{ij} $s are given in \ref{app:asymptote}. The asymptotic form up to second order is given by
	\begin{align}
	\begin{split}
		&\begin{pmatrix} f_i^{(\text{finite }\epsilon)} \\ g_i^{(\text{finite }\epsilon)} \end{pmatrix} \rightarrow \begin{pmatrix} \delta_{i1} \\ 0 \end{pmatrix} + \mathrm{i}\epsilon \begin{pmatrix} l_{i1}x+l_{i0}\operatorname{sgn}x \\ m_{i1} \end{pmatrix} \\ 
		&\quad-\frac{\epsilon^2}{2\kappa^2}\begin{pmatrix}\frac{1}{\rho_\infty}\Bigl[ (p_{\text{L}}^2m_{i1}-jl_{i1})\frac{x^2}{2}+(p_{\text{L}}^2m_{i0}-jl_{i0})|x|\Bigr]\!+\!c_{i0}\! \\ (\rho_\infty l_{i1}-pm_{i1})x+(\rho_\infty l_{i0}-pm_{i0})\operatorname{sgn}x \end{pmatrix}+O(\epsilon^3).
	\end{split}
	\end{align}
	Here $ c_{i0} $ is a certain constant whose form is not important here.\\ 
	\indent Let  $ (u^{(\text{finite }\epsilon)}_i,v^{(\text{finite }\epsilon)}_i) $ be the counterpart of  $ (f^{(\text{finite }\epsilon)}_i,g^{(\text{finite }\epsilon)}_i) $ in  $ (u,v) $ notation. If we write them in plane wave form
	\begin{align}
		\begin{pmatrix} u^{(\text{finite }\epsilon)}_1 \\ v^{(\text{finite }\epsilon)}_1 \end{pmatrix} &\rightarrow \frac{\mathrm{i}\sqrt{\rho_\infty}}{2pp_{\text{L}}}\left[ C_1^{\pm}w_1\left( x,\pm\tfrac{\delta}{2} \right)+C_2^{\pm}w_2\left( x,\pm\tfrac{\delta}{2} \right) \right] \quad (x\rightarrow\pm\infty) \\
		\intertext{and}
		\begin{pmatrix} u^{(\text{finite }\epsilon)}_3 \\ v^{(\text{finite }\epsilon)}_3 \end{pmatrix} &\rightarrow \frac{\mathrm{i}\sqrt{\rho_\infty}}{2pp_{\text{L}}}\left[ D_1^{\pm}w_1\left( x,\pm\tfrac{\delta}{2} \right)+D_2^{\pm}w_2\left( x,\pm\tfrac{\delta}{2} \right) \right] \quad (x\rightarrow\pm\infty),
	\end{align}
	then the energy dependence of coefficients can be obtained as follows:
	\begin{align}
		C_1^\pm &= p(p+p_{\text{L}}+2\kappa^2l_{11})\pm\mathrm{i}\epsilon p_{\text{L}}(p-p_{\text{L}})l_{10}+O(\epsilon^2), \label{eq:C1} \\
		C_2^\pm &= p(-p+p_{\text{L}}-2\kappa^2l_{11})\pm\mathrm{i}\epsilon p_{\text{L}}(p+p_{\text{L}})l_{10}+O(\epsilon^2), \\
		D_1^\pm &= 2p\kappa^2 l_{31}\pm\mathrm{i}\epsilon p_{\text{L}}\left( \frac{p_{\text{L}}N}{4\rho_\infty}+(p-p_{\text{L}})l_{30} \right)+O(\epsilon^2), \\
		D_2^\pm &= -2p\kappa^2 l_{31}\pm\mathrm{i}\epsilon p_{\text{L}}\left( -\frac{p_{\text{L}}N}{4\rho_\infty}+(p+p_{\text{L}})l_{30} \right)+O(\epsilon^2). \label{eq:D2}
	\end{align}
	The striking feature is that the zeroth order of  $ D_1^\pm $ and $ D_2^\pm $ vanishes when the soliton velocity becomes the critical value, see Eq. (\ref{eq:l31}). This is an immediate cause of the singular behavior of the reflection coefficient.\\
	\indent A solution of the scattering problem (\ref{eq:uvasymptotic}) is constructed as follows:
	\begin{align}
		\begin{pmatrix}u \\ v \end{pmatrix}=D_2^+\begin{pmatrix} u^{(\text{finite }\epsilon)}_1 \\ v^{(\text{finite }\epsilon)}_1 \end{pmatrix}-C_2^+\begin{pmatrix} u^{(\text{finite }\epsilon)}_3 \\ v^{(\text{finite }\epsilon)}_3 \end{pmatrix},
	\end{align}
	and coefficients $ t $ and $ r $ are given by
	\begin{align}
		t&=\frac{C_2^+D_1^+-D_2^+C_1^+}{C_2^+D_1^--D_2^+C_1^-}, \\
		r&=\frac{C_2^+D_2^--D_2^+C_2^-}{C_2^+D_1^--D_2^+C_1^-}.
	\end{align}
	Finally, moving to the particular coordinate system $ (\alpha,\beta)=(p,\rho_\infty) $, and using the expressions given in \ref{app:asymptote}, we get the main result
	\begin{align}
		-\frac{2\rho_\infty^2\delta_p}{p^2p_{\text{L}}^2}(C_2^+D_1^--D_2^+C_1^-) &= a P_p-\mathrm{i}\epsilon\left( b+b_1P_p \right)+O(\epsilon^2) \label{eq:final1}\\
		\intertext{and}
		-\frac{2\rho_\infty^2\delta_p}{p^2p_{\text{L}}^2}(C_2^+D_2^--D_2^+C_2^-) &= -\mathrm{i}\epsilon\left( d+d_1P_p \right)+O(\epsilon^2) \label{eq:final2}
	\end{align}
	with
	\begin{align}
		a &= 4p_{\text{L}}\rho_\infty, \\
		b &= (N+pN_p)^2+(p_{\text{L}}N_p)^2, \\
		b_1 &= N-\left( 1+\frac{2p^2}{\kappa^2} \right)[N,j]_{p\rho_\infty}, \\
		d &= (N+pN_p)^2-(p_{\text{L}}N_p)^2, \\
		d_1 &= N+[N,j]_{p\rho_\infty}. \label{eq:final7}
	\end{align}
	It gives the result in Subsec. \ref{subsec:mainresult}.
\section{Discussions and Concluding Remarks}\label{sec:summary}
	\indent In this last section, we give a few discussions and future perspectives.
\subsection{Local density fluctuation at the critical point}
	In the system with a potential wall, the emergence of a zero-energy local density fluctuation was a key of the destabilization of superflow\cite{TakahashiKato, TakahashiKatoConf}. In the present case of solitons, an amplification of the zero-energy local density fluctuation also occurs at the critical point, but its mathematical structure slightly differs. Let us see it in detail. \\ 
	\indent Before beginning, one should recall that $ f $ and $ g $ have the meaning of phase and density fluctuations. (See subsection \ref{subsec:fg} again.) So, the solution $ (u,v)\propto(\phi,\phi^*) \,\leftrightarrow\, (f,g)\propto(1,0) $ has no density fluctuation. If another solution, e.g., a parameter-derivative solution, is added, a non-zero density fluctuation arises. \\
	\indent For a superflow state against the potential barrier\cite{BaratoffBlackburnSchwartz,Hakim,PhamBrachet}, it is known\cite{TakahashiKato,TakahashiKatoConf} that
	\begin{align}
		\lim_{\epsilon\rightarrow0}\begin{pmatrix} u \\ v \end{pmatrix} &= \begin{pmatrix} \phi \\ \phi^* \end{pmatrix} & \text{(for non-critical states)}. \\ 
		\lim_{\epsilon\rightarrow0}\begin{pmatrix} u \\ v \end{pmatrix} &= \begin{pmatrix} \phi \\ \phi^* \end{pmatrix}+c \dfrac{\partial }{\partial \varphi}\begin{pmatrix} \phi \\ -\phi^* \end{pmatrix} & \text{(for a critical state)}.
	\end{align}
	Here $ \varphi $ is a Josephson phase difference and  $ c $ is a certain constant. Thus, the density fluctuation represents the anomaly of the critical point. \\
	\indent In the case of solitons, however, because of spontaneous translational symmetry breaking, the density fluctuating zero-mode  $ (f_3,g_3) $ (or equivalently, $ (u_3,v_3) $ )  always exists, and this mode indeed contributes to the solution of the scattering problem:
	\begin{align}
		\lim_{\epsilon\rightarrow0}\begin{pmatrix} u \\ v \end{pmatrix} = c_1 \begin{pmatrix} u_1 \\ v_1 \end{pmatrix} +c_3 \begin{pmatrix} u_3 \\ v_3 \end{pmatrix} \quad \text{(for non critical states).}
	\end{align}
	Here  $ c_1=\lim_{\epsilon\rightarrow0}(-C_2^+) $ and $ c_3=\lim_{\epsilon\rightarrow0}D_2^+ $ are constants. (See Subsec.~\ref{subsec:2ndasym} for more detailed expressions.)  Therefore, the local density fluctuation always exists regardless of whether the soliton is stable or unstable. However, when the soliton velocity comes closer to the critical one, the ratio $ c_3/c_1 $ becomes larger, and it becomes infinite at the critical point. That is to say, at the critical velocity state, 
	\begin{align}
		\lim_{\epsilon\rightarrow0}\begin{pmatrix} u \\ v \end{pmatrix} = \begin{pmatrix} u_3 \\ v_3 \end{pmatrix} \quad \text{(for a critical state)}
	\end{align}
	holds. Thus, we can say that the amplification of the local density fluctuation also plays a key role in the case of the destabilization of solitons.
\subsection{Conclusions and Future perspectives}
	In this paper, we have solved the scattering problem of linearized excitations (Bogoliubov phonons) against a dark soliton in a generalized NLS system. We have exactly shown that the perfect transmission of a zero-energy phonon vanishes when the soliton velocity reaches the critical value, and near the critical velocity state, the reflection coefficient obeys a saddle-node type universal scaling law. Our result has a fundamental importance because it provides an exact example of saddle-node scaling in infinite dimensional time-reversible Hamiltonian systems. Through the proof, we have also obtained the exact zero-energy solutions and their finite energy generalizations. In the derivation of them, the use of two kinds of parameter derivatives has played an important role. This method will also be useful to elucidate the low-energy physics of other systems, such as higher dimensional systems or multi-component systems. \\
	\indent Even though we have shown the example of scaling laws, the derivation of a normal form of saddle-node bifurcation remains unsolved. The similar problem for the supercurrent-flowing system in the presence of an obstacle\cite{Hakim,PhamBrachet} also exists. Compared to the problem of solitons treated in this paper, the system with an obstacle seems to be a little more difficult, because all four zero-energy solutions for linearized equation are not yet obtained except for the critical velocity state\cite{TakahashiKato}. These issues are left as future works. \\
	
	\indent In this paper we have treated the soliton-phonon scattering problem. One interesting generalization is the multi-soliton scattering process. It was shown that the existence of the two solitons makes the reflection coefficient of phonons non-trivial even for the integrable cubic case\cite{KolbyshovaSadreev}. The soliton collision problem in the non-integrable generalized KdV equation was recently investigated\cite{MartelMerleInventMath,MartelMerleAnnMath}. The similar problem of dark solitons in non-integrable NLS systems is also important to understand the generic solitary characters of dark solitons.\\
	
	\indent An emergence of singularity of scattering properties at a critical state which separates the stable and unstable branches is expected to be a universal character in more general systems, because the emergent or amplified zero-modes can affect the transmission properties of low-energy modes. Since the scattering problem of linearized excitations is easier and more analytically tractable than the existence-proof of an unstable mode or construction of a Lyapunov function, it will be useful to ``conjecture'' the criterion for stability of solitons, even though it is not a direct proof of the stability itself. For example, to the best knowledge of the present author, the stability criterion of dark solitons in multi-component NLS systems is an untouched problem. In ultracold atoms, the binary mixture of Rb atoms is realized\cite{HoShenoy,Myatt} and it is known that the corresponding coupled NLS equation becomes integrable only for the Manakov case. Also, the Bose condensates with spin degree of freedom are created by optical trap\cite{StamperKurn}, and Wadati group members have studied the spin-1 solitons at the integrable point\cite{TsuchidaWadati,IedaMiyakawaWadati,UchiyamaIedaWadati}. When the coupling constant deviates from the integrable one, it is no longer ensured that a dark soliton is always stable. So, it may be an interesting problem to study the stability of solitons through the study of scattering problems of linearized excitations.\\ 
	
\paragraph{Acknowledgment}
	The author is grateful to Y. Kato and M. Kunimi for helpful comments. This work was supported by a Grant-in-Aid for JSPS Fellows (No. 22-10058).

\appendix

\section{Identities for parameter derivatives}\label{app:idnty}
	Here we derive a few identities for parameter derivatives. The NLS equation (\ref{eq:nls2}) expressed in terms of the density $ \rho(x) $ is given by
	\begin{align}
		\rho_{xx}=2\rho\left( \frac{\rho_x^2}{4\rho^2}+\frac{j^2}{\rho^2}-\mu+F(\rho) \right).
	\end{align}
	From the parameter derivative of Eq. (\ref{eq:momconsrv}) and the above equation, one obtains
	\begin{align}
		\rho_\alpha\rho_{xx}-\rho_{\alpha x}\rho_x = 2(\rho_\infty-\rho)(2p j_\alpha-\mu_\alpha \rho). \label{eq:syspar}
	\end{align}
	Here recall that $ j=\rho_\infty p $ and $ \mu = p^2+F(\rho_\infty) $. The same expression follows by replacement $ \alpha\rightarrow\beta $. From them, an important identity 
	\begin{align}
		g_3g_2'-g_2g_3' = -4p(\rho_\infty-\rho) \label{eq:wsimplify}
	\end{align}
	follows, where $ g_2 $ and $ g_3 $ are defined in Subsec.~\ref{subsec:zerosol}. \\
	\indent Dividing both sides of (\ref{eq:syspar}) by $ \rho $, and integrating them from $ -\infty $ to $ +\infty $, we obtain
	\begin{align}
		2j_\alpha\delta+\mu_\alpha N=-\frac{\partial }{\partial \alpha}\int_{-\infty}^\infty\frac{(\rho_x)^2\mathrm{d}x}{2\rho}.
	\end{align}
	Since the same expression for $ \beta $ holds, we obtain the second important identity
	\begin{align}
	\begin{split}
		& (2j_\alpha\delta+\mu_\alpha N)_\beta=(2j_\beta\delta+\mu_\beta N)_\alpha \\
		\leftrightarrow\quad & 2[j,\delta]_{\alpha\beta}+[\mu,N]_{\alpha\beta}=0. \label{eq:idntydev}
	\end{split}
	\end{align}
\section{Derivation of the fourth zero-energy solution}\label{app:roo}
	Let us derive the expression for the fourth zero-energy solution $ (f_4,g_4) $. Eliminating $ f $ from Eqs. (\ref{eq:bogofg1}) and (\ref{eq:bogofg2}), one obtains the third order differential equation for $ g $:
	\begin{align}
		g'''+A g'+ B=0.
	\end{align} 
	Here $ A $ and $ B $ are some functions, but their forms are not important here. Knowing two solutions $ g_2 $ and $ g_3 $, the last one is obtained by reduction of order:
	\begin{align}
		\tilde{g}_4 &:= g_2 \int_0^x\frac{g_3\mathrm{d}x}{w^2}-g_3\int_0^x\frac{g_2\mathrm{d}x}{w^2}
		\intertext{with}
		w &= g^{}_3g_2'-g^{}_2g_3'.
	\end{align}
	Since $ w $ is given by (\ref{eq:wsimplify}), the above expression is simplified as
	\begin{align}
		16p^2\tilde{g}_4 = \left[ \frac{1}{\rho_\infty-\rho}-\frac{1}{\rho_\infty-\rho_0} \right]g_2- g_3\int_0^x\frac{g_2\mathrm{d}x}{(\rho_\infty-\rho)^2}
	\end{align}
	Here $ \rho_0:=\rho(0) $. The solution which cancels out the constant (\ref{eq:zeroBogoconst}) is given by the following linear combination with $ g_2 $:
	\begin{align}
		g_4:=-\frac{\rho_0}{\rho_\infty-\rho_0}g_2-16p^2\rho_\infty \tilde{g}_4,
	\end{align} 
	and  $ f_4 $ is calculated from $ f_4' = -jg_4/\rho^2 $. 
\section{Asymptotics of zero-energy solutions}\label{sec:asymofzero}
	Let us write the asymptotic form of  $ \rho(x) $ as
	\begin{align}
		\rho(x) = \rho_\infty - \rho_\infty^{(1)}\mathrm{e}^{-2\kappa|x|}+\dotsb,
	\end{align}
	then a half of the healing length $ \kappa $ is given by
	\begin{align}
		\kappa=\sqrt{p_{\text{L}}^2-p^2}.
	\end{align}
	From (\ref{eq:nlsdsasym}) or (\ref{eq:phasecond}), an asymptotic form of the phase  $ S(x) $ is given by 
	\begin{align}
		S(x) \rightarrow px+(\operatorname{sgn}x)\frac{\delta}{2}.
	\end{align}
	Using the invariant nature of the ratio (\ref{eq:invratio}), one can show
	\begin{align}
		\frac{[\mu,p]_{\alpha\beta}}{[\mu,j]_{\alpha\beta}} &= \frac{[\mu,p]_{p\rho_\infty}}{[\mu,j]_{p\rho_\infty}} = \frac{p_{\text{L}}^2}{\rho_\infty \kappa^2}, \\
		\frac{[\mu,\rho_\infty]_{\alpha\beta}}{[\mu,j]_{\alpha\beta}} &= \frac{[\mu,\rho_\infty]_{p\rho_\infty}}{[\mu,j]_{p\rho_\infty}} = -\frac{p}{\kappa^2}.
	\end{align}
	With the use of the above relations, the asymptotic forms of zero energy solutions are evaluated as follows: 
	\begin{align}
		f_2 & \rightarrow \frac{p_{\text{L}}^2}{\rho_\infty \kappa^2}x+(\operatorname{sgn}x)\frac{1}{2}\frac{[\mu,\delta]_{\alpha\beta}}{[\mu,j]_{\alpha\beta}}, \\
		g_2 & \rightarrow -\frac{p}{\kappa^2}, \\
		f_3 & \rightarrow \frac{p\rho_\infty^{(1)}}{\rho_\infty}\mathrm{e}^{-2\kappa|x|}, \\
		g_3 & \rightarrow (\operatorname{sgn}x) 2\kappa \rho_\infty^{(1)}\mathrm{e}^{-2\kappa|x|}, \\
		f_4 & \rightarrow -(\operatorname{sgn}x)\frac{p^2}{4\rho_\infty^{(1)}\kappa^3}\mathrm{e}^{2\kappa|x|}, \\
		g_4 & \rightarrow \frac{j}{2\rho_\infty^{(1)}\kappa^2}\mathrm{e}^{2\kappa|x|}.
	\end{align}
\section{Symmetry consideration on zero-modes}\label{app:symmetry}
	In this appendix we consider how a symmetry plays a role in finding a solution of a linearized equation. Particularly, we emphasize that the information we can obtain from a Galilean symmetry is not about the zero energy solution but about the first order correction of a finite energy solution.\\ 
	\indent Let  $ \phi(x,t) $  be a solution of time-dependent NLS equation (\ref{eq:nls}) and $ \tilde{\phi}(x,t,\alpha) $ be a family of solutions with continuous parameter $ \alpha $ such that $ \tilde{\phi}(x,t,0)=\phi(x,t) $.  Here we assume that  $ \alpha $ is free from any system parameter. (The parameter derivative introduced in Subsec. \ref{subsec:paradera} is a more general concept, because it can depend on the system parameters which appear, e.g.,  in a boundary condition (\ref{eq:nlsdsasym}).) Differentiation of NLS equation (\ref{eq:nls}) with respect to $ \alpha $ immediately yields
	\begin{align}
		(\mathrm{i}\partial_t-\mathcal{L})\begin{pmatrix} \phi_\alpha \\ -\phi^*_\alpha \end{pmatrix} = 0 
	\end{align} 
	with a definition $ \phi_\alpha:= [\partial_\alpha\tilde{\phi}]_{\alpha=0} $. Thus, $ \phi_\alpha $ is a solution of a \textit{time-dependent} linearized equation (\ref{eq:tdbogo}) in the presence of the condensate wavefunction $ \phi(x,t) $. Particularly, if one sets $ \tilde{\phi}(x,t,\alpha)=\phi(x,t)\mathrm{e}^{\mathrm{i}\alpha} $, $ \tilde{\phi}(x,t,\alpha)=\phi(x+\alpha,t) $, and Eq. (\ref{eq:galilei}), which represent a global phase symmetry, a translational symmetry, and a Galilean symmetry, respectively, one obtains the following particular solutions:
	\begin{align}
		(\mathrm{i}\partial_t-\mathcal{L})\begin{pmatrix} \mathrm{i}\phi \\ \mathrm{i}\phi^* \end{pmatrix} &= 0, \\
		(\mathrm{i}\partial_t-\mathcal{L})\begin{pmatrix} \phi_x \\ -\phi^*_x \end{pmatrix} &= 0, \\
		(\mathrm{i}\partial_t-\mathcal{L})\begin{pmatrix} 2t\phi_x-\mathrm{i}x\phi \\ -2t\phi^*_x-\mathrm{i}x\phi^* \end{pmatrix} &= 0.
	\end{align}
	It is a result for a time-dependent equation. In order to interpret these results to a \textit{stationary} problem, let us set $ \phi(x,t)=\phi(x)\mathrm{e}^{-\mathrm{i}\mu t} $. From the assumption stated above,  $ \mu $ does not depend on $ \alpha $. (On the other hand,  $ \mu $ can depend on the parameter in Subsec. \ref{subsec:paradera}.) The above equations are rewritten as
	\begin{align}
		\mathcal{L}_\mu\begin{pmatrix} \mathrm{i}\phi \\ \mathrm{i}\phi^* \end{pmatrix} &= 0, \label{eq:appsymzero1} \\
		\mathcal{L}_\mu\begin{pmatrix} \phi_x \\ -\phi^*_x \end{pmatrix} &= 0, \label{eq:appsymzero2} \\
		2\mathrm{i}\begin{pmatrix} \phi_x \\ -\phi^*_x \end{pmatrix}-2t\mathcal{L}_\mu\begin{pmatrix} \phi_x \\ -\phi^*_x \end{pmatrix}+\mathcal{L}_\mu\begin{pmatrix} \mathrm{i}x\phi \\ \mathrm{i}x\phi^* \end{pmatrix} &=0. \label{eq:appsymzero3}
	\end{align}
	Here  $ \mathcal{L}_\mu $ defined by Eqs. (\ref{eq:bogos}) and (\ref{eq:bogos2}) is a differential operator of the stationary Bogoliubov equation.  (\ref{eq:appsymzero1}) and (\ref{eq:appsymzero2}) immediately give zero-energy solutions, whereas we need a consideration on Eq. (\ref{eq:appsymzero3}). Since this equation must hold for any time $ t $, both coefficients of $ t^0 $ and $ t^1 $ must vanish. The equation for the $ t^1 $-coefficient reproduces (\ref{eq:appsymzero2}). From the $ t^0 $-coefficient, one obtains 
	\begin{align}
		-\frac{1}{2}\mathcal{L}_\mu\begin{pmatrix} x\phi \\ x\phi^* \end{pmatrix} = \begin{pmatrix} \phi_x \\ -\phi^*_x \end{pmatrix}.
	\end{align}
	It represents Eq. (\ref{eq:expansion2}) with $ n=1 $ and $ (u^{(0)},v^{(0)})=(\phi_x,-\phi^*_x) $. Therefore, it gives the first order solution of (\ref{eq:1stu3v3}). Thus, a Galilean symmetry gives us a piece of information about first order solutions, and gives no new information about zero-energy solutions. A linearly divergent zero-energy solution, which exists even when the system does not have a translational or a Galilean symmetry, can be obtained by two kinds of parameter-derivatives stated in Subsec. \ref{subsec:paradera}. \\
	\indent We further note that Eq. (\ref{eq:1stu3v3}) itself does not describe a physically meaningful non-divergent solution. As we show in Subsec. \ref{subsec:boundedsols}, we must construct a linear combination of solutions so that the second order term is free from exponential divergence. A correct first order solution with non-divergent character is given by (\ref{eq:f31g31}).
\section{Formulae for calculation of asymptotes}\label{app:asymptote}
	Let us derive some formulae for  $ l_{ij} $s and $ m_{ij} $s defined by Eqs. (\ref{eq:f1asym}) and (\ref{eq:g1asym}). Using the relations 
	\begin{align}
		[X,j]_{\alpha\beta} &= \rho_\infty[X,p]_{\alpha\beta}+p[X,\rho_\infty]_{\alpha\beta}, \label{eq:xj}\\
		[X,\mu]_{\alpha\beta} &= 2p[X,p]_{\alpha\beta}+F'(\rho_\infty)[X,\rho_\infty]_{\alpha\beta}, \label{eq:xmu}
	\end{align}
	one can show
	\begin{align}
		\rho_\infty[X,\mu]_{\alpha\beta}-2p[X,j]_{\alpha\beta}&=2\kappa^2[X,\rho_\infty]_{\alpha\beta}. \label{eq:xjxmu}
	\end{align}
	Here remember that $ p_{\text{L}} $ and $ F'(\rho_\infty) $ are related to each other by (\ref{eq:defofpl}), and $ \kappa^2=p_{\text{L}}^2-p^2 $. Using it and (\ref{eq:idntydev}), one obtains
	\begin{align}
		\rho_\infty[\delta,\mu]_{\alpha\beta}+p[N,\mu]_{\alpha\beta}&=2\kappa^2[\delta,\rho_\infty]_{\alpha\beta}. \label{eq:deltamu}
	\end{align}
	From (\ref{eq:f11g11}), (\ref{eq:f31g31}), and (\ref{eq:deltamu}), we can obtain
	\begin{align}
		l_{11} &= \frac{\rho_\infty[\delta,p]_{\alpha\beta}+p[N,p]_{\alpha\beta}}{2\kappa^2[\delta,\rho_\infty]_{\alpha\beta}}, \\
		m_{11} &= \frac{\rho_\infty[\delta,\rho_\infty]_{\alpha\beta}+p[N,\rho_\infty]_{\alpha\beta}}{2\kappa^2[\delta,\rho_\infty]_{\alpha\beta}}, \\
		l_{10} &= \frac{p[N,\delta]_{\alpha\beta}}{4\kappa^2[\delta,\rho_\infty]_{\alpha\beta}}, \\
		m_{10} &= \frac{\rho_\infty[\delta,N]_{\alpha\beta}}{4\kappa^2[\delta,\rho_\infty]_{\alpha\beta}}, \\
		l_{31} &= \frac{1}{2}-\frac{N[\mu,p]_{\alpha\beta}}{4\kappa^2[\delta,\rho_\infty]_{\alpha\beta}}-pl_{11}, \\
		m_{31} &= -\frac{N[\mu,\rho_\infty]_{\alpha\beta}}{4\kappa^2[\delta,\rho_\infty]_{\alpha\beta}}-pm_{11}, \\
		l_{30} &= -\frac{N[\mu,\delta]_{\alpha\beta}}{8\kappa^2[\delta,\rho_\infty]_{\alpha\beta}}-pl_{10}, \\
		m_{30} &= -\frac{N[\mu,N]_{\alpha\beta}}{8\kappa^2[\delta,\rho_\infty]_{\alpha\beta}}-pm_{10}.
	\end{align}
	Further, using them and (\ref{eq:xj}), (\ref{eq:xmu}), and (\ref{eq:xjxmu}),
	\begin{align}
		2p l_{11}+F'(\rho_\infty)m_{11} &= 1, \label{eq:l11m11} \\
		\rho_\infty l_{11}+pm_{11}&=-\frac{1}{2}\frac{[N,\rho_\infty]_{\alpha\beta}}{[\delta,\rho_\infty]_{\alpha\beta}}, \\
		2p l_{10}+F'(\rho_\infty)m_{10} &= -\frac{1}{2}\frac{[N,\delta]_{\alpha\beta}}{[\delta,\rho_\infty]_{\alpha\beta}}, \\
		\rho_\infty l_{10}+pm_{10}&=0, \label{eq:l10m10} \\
		2p l_{31}+F'(\rho_\infty)m_{31} &= 0, \label{eq:l31m31}\\
		\rho_\infty l_{31}+pm_{31}&=-\frac{1}{2}\frac{[P,\rho_\infty]_{\alpha\beta}}{[\delta,\rho_\infty]_{\alpha\beta}}, \\
		2p l_{30}+F'(\rho_\infty)m_{30} &=\frac{1}{2}\frac{[pN,\delta]_{\alpha\beta}}{[\delta,\rho_\infty]_{\alpha\beta}}, \\
		\rho_\infty l_{30}+pm_{30}&=\frac{N}{4}. \label{eq:l30m30}
	\end{align}
	We can eliminate $ m_{ij} $s by Eqs. (\ref{eq:l11m11}), (\ref{eq:l10m10}), (\ref{eq:l31m31}), and (\ref{eq:l30m30}), and the expressions (\ref{eq:C1})--(\ref{eq:D2}) are derived in that way.\\ 
	\indent Now we consider the particular coordinate  $ (\alpha,\beta)=(p,\rho_\infty) $.  $ l_{ij} $s are then given by
	\begin{align}
		l_{11}&=-\frac{j\delta_p+p_{\text{L}}^2N_p}{2\rho_\infty\kappa^2\delta_p}, \label{eq:l11} \\
		l_{10}&=\frac{p[N,\delta]_{p\rho_\infty}}{4\kappa^2\delta_p}, \\
		l_{31}&=-\frac{p_{\text{L}}^2 P_p}{2\rho_\infty\kappa^2\delta_p}, \label{eq:l31}\\
		l_{30}&=-\frac{N[\mu,\delta]_{p\rho_\infty}}{8\kappa^2\delta_p}-pl_{10}.
	\end{align}
	From  $ P = -pN-\rho_\infty\delta $ and (\ref{eq:idntydev}), we obtain
	\begin{align}
		\delta_p &= -\frac{N+pN_p+P_p}{\rho_\infty}, \\
		\delta_{\rho_\infty} &= -\frac{pN-\kappa^2N_p+pP_p+jN_{\rho_\infty}}{\rho_\infty^2}. \label{eq:deltarhoinfty}
	\end{align}
	Derivatives of $ \delta $ are eliminated by using them. With the use of Eqs. (\ref{eq:l11})--(\ref{eq:deltarhoinfty}), we can obtain the final result (\ref{eq:final1})--(\ref{eq:final7}).
\section{Zero-energy solutions without using parameter derivatives}\label{app:zerosol}
	We can write down the general zero-energy solution without using parameter derivatives:
	\begin{align}
		\begin{pmatrix}u\\ v \end{pmatrix}&=\sum_{i=1,2,3,4} c_i \begin{pmatrix}u_i^{} \\ -u_i^* \end{pmatrix}, \\
		u_1 &= \mathrm{i}\phi, \\
		u_2 &= \phi_x, \\
		u_3 &= \mathrm{i}\phi\int\!\frac{\mathrm{d}x}{\rho}+4\mathrm{i}j^2\phi\int\!\frac{\mathrm{d}x}{\rho\rho_x^2}-4j\phi_x\int\!\frac{\mathrm{d}x}{\rho_x^2}, \\
		u_4 &= \mathrm{i}j\phi\int\!\frac{\mathrm{d}x}{\rho_x^2}-\phi_x\int\!\frac{\rho\mathrm{d}x}{\rho_x^2}.
	\end{align}
	These expressions, however, have a fatal flaw; Namely, they have artificial singularities at the origin (or more precisely, the points where $ \rho'(x)=0 $.) It means that these expressions do not give \textit{global} solutions which continuously connect the solutions from $ x=-\infty $ to $ x=+\infty $, instead, only give  \textit{local} solutions which satisfy the differential equation at each point. For this reason they are not so useful in scattering problems, which are equivalent to the determination of global behavior of given solutions.  On the other hand, the parameter-derivative solutions given in Subsec. \ref{subsec:zerosol} have no singularities if the density $ \rho(x) $ does not cross the value $ \rho_\infty $. 
\section{Stationary solutions of CQNLS equation}\label{app:cqnls}
	In the CQNLS system, the equation for momentum conservation (\ref{eq:genjm}) becomes
	\begin{align}
		\frac{(\rho_x)^2}{4}=-j^2+j_m\rho-\mu\rho^2+a_1\rho^3+a_2\rho^4. \label{eq:cqdiff}
	\end{align}
	Here we want general stationary solutions, so we do not concentrate on the solution with the asymptotic form (\ref{eq:nlsdsasym}), and therefore $ \mu, j, $ and $ j_m $ need not be given by  (\ref{eq:cp}), (\ref{eq:dsj}), and (\ref{eq:dsjm}), respectively.  If the right hand side of (\ref{eq:cqdiff}) is factored as  $ a_2\prod_{i=1}^4(\rho-\rho_i) $, the solution of this differential equation is given by the following cross-ratio form:
	\begin{align}
		\operatorname{sn}^2\left( \kappa x | m \right)&=\frac{(\rho(x)-\rho_1)(\rho_3-\rho_2)}{(\rho(x)-\rho_2)(\rho_3-\rho_1)} \\
		\intertext{with}
		\kappa&=\sqrt{a_2(\rho_4-\rho_2)(\rho_3-\rho_1)}, \label{eq:kappaapp} \\
		m&=\frac{(\rho_4-\rho_1)(\rho_3-\rho_2)}{(\rho_4-\rho_2)(\rho_3-\rho_1)},
	\end{align}
	or equivalently,
	\begin{align}
		\rho(x)=\frac{\rho_2(\rho_3-\rho_1)-\rho_1(\rho_3-\rho_2)\operatorname{sn}^2\left( \kappa x | m \right)}{(\rho_3-\rho_1)-(\rho_3-\rho_2)\operatorname{sn}^2\left( \kappa x | m \right)}. \label{eq:rhoapp}
	\end{align}
	Here we use Mathematica's definition for Jacobi elliptic functions.\footnote{The Wolfram Functions Site, http://functions.wolfram.com/} \\
	\indent For the case of the dark soliton solution,  $ \rho_3=\rho_4=\rho_\infty $ holds, so $ m=1 $ follows and the sn function reduces to the tanh function. $ \rho_1 $ and $ \rho_0:=\rho_2 $ are given by the roots of
	\begin{align}
		a_2\rho^2+(2a_2\rho_\infty+a_1)\rho-p^2=0. \label{eq:quadracqnls}
	\end{align}
	$ \rho_0 $ (\ref{eq:rho0cqnls}) is the root of Eq. (\ref{eq:quadracqnls}) with a plus sign.  $ \rho_1 $ can be eliminated in two ways; From Eq. (\ref{eq:kappaapp}) or from the fact that $ \rho_0 $ and $ \rho_1 $ solve Eq. (\ref{eq:quadracqnls}),
	\begin{align}
		\rho_\infty-\rho_1=\frac{\kappa^2}{a_2(\rho_\infty-\rho_0)}, \qquad \rho_0\rho_1=-\frac{p^2}{a_2}. \label{eq:eliminaterho1}
	\end{align}
	The expression (\ref{eq:rhoapp}) is then rewritten as
	\begin{align}
		\rho(x)=\rho_\infty-\frac{\kappa^2(\rho_\infty-\rho_0)\operatorname{sech}^2\kappa x}{\kappa^2-a_2(\rho_\infty-\rho_0)^2\tanh^2\kappa x}.
	\end{align}
	The following expressions are also useful for calculation of the phase shift $ \delta $ and the particle number of soliton $ N $: 
	\begin{align}
		S_x=\frac{j}{\rho}&= p+\frac{(\rho_0\kappa)[ p(\rho_\infty-\rho_0)\tanh\kappa x]'}{(\rho_0\kappa)^2+[ p(\rho_\infty-\rho_0)\tanh\kappa x]^2}, \label{eq:Sxinapp} \\
		\rho-\rho_\infty &= -\frac{1}{\sqrt{a_2}}\frac{\kappa [\!\sqrt{a_2}(\rho_\infty-\rho_0)\tanh\kappa x]'}{\kappa^2-[\!\sqrt{a_2}(\rho_\infty-\rho_0)\tanh\kappa x ]^2},
	\end{align}
	where Eq. (\ref{eq:Sxinapp}) is obtained by eliminating both $ a_2 $ and  $ \rho_1 $ with the use of (\ref{eq:eliminaterho1}). Here we can use the formulae
	\begin{align}
		\frac{ay'}{a^2+y^2} &= \Bigl(\tan^{-1}\frac{y}{a}\Bigr)', \\
		\frac{ay'}{a^2-y^2} &= \Bigl(\tanh^{-1}\frac{y}{a}\Bigr)'.
	\end{align}
	Thus the content in Subsec.~\ref{subsec:cqnls} is reproduced. \\



\bibliographystyle{model1-num-names}
\bibliography{genNLS}







\end{document}